\documentclass[12pt,preprint]{aastex}

\usepackage{graphicx, amsmath, amssymb, amsthm}

\makeatother

\newcommand\etal{{\it et~al.~}}

\newcommand\Ms{M_\odot}

\begin{document}
\title{How the First Stars Regulated Local Star Formation I: Radiative Feedback}
\author{Daniel Whalen\altaffilmark{1}, Brian W. O'Shea\altaffilmark{1,3},
Joseph Smidt\altaffilmark{2} \& Michael L. Norman \altaffilmark{4}}
\altaffiltext{1}{Applied Physics (X-2), Los Alamos National
Laboratory. Email: dwhalen@lanl.gov}
\altaffiltext{2}{Department of Physics and Astronomy, Brigham Young University, Provo,
UT 84602}
\altaffiltext{3}{Theoretical Astrophysics (T-6), Los Alamos National
Laboratory}
\altaffiltext{4}{Center for Astrophysics and Space Sciences,
University of California at San Diego, La Jolla, CA 92093, U.S.A.}

\begin{abstract} 

We present numerical simulations of how a 120 M$_\odot$ primordial star regulates star formation in 
nearby cosmological halos at $z \sim$ 20 by photoevaporation.  Our models include nine-species 
primordial chemistry and self-consistent multifrequency conservative transfer of UV photons
with all relevant radiative processes.  Whether or not new stars form in halos clustered around a
Population III star ultimately depends on their core densities and proximity to the star.  Diffuse 
halos with central densities below 2 - 3 cm$^{-3}$ are completely ionized and evaporated anywhere 
in the cluster.  Evolved halos with core densities above 2000 cm$^{-3}$ are impervious to both 
ionizing and Lyman-Werner flux at most distances from the star, and collapse as quickly as they
would in its absence.  Star formation in halos of intermediate density can be either promoted 
or suppressed depending on how the I-front remnant shock compresses, deforms and enriches the 
core with H$_2$.  We find that the 120 M$_\odot$ star photodissociates H$_2$ in most halos in 
the cluster but that catalysis by H- restores it a few hundred kyr after the death of the star, 
with little effect on star formation.  Our models exhibit significant departures from previous 
one-dimensional spherically-symmetric simulations, which are prone to serious errors due to 
unphysical geometric focusing effects.

\end{abstract}


\section{Introduction}

Radiative feedback of one generation of stars on the next has regulated the rise of
stellar populations since the birth of the first luminous objects in the universe.  In the 
$\Lambda$CDM paradigm of hierarchical structure formation, small dark matter halos at high
redshifts assembled into larger structures by collisions and mergers.  When
the first pregalactic objects reached masses of $\sim$ 1.0 $\times$ 10$^5$ $\Ms$ at $z \sim$ 
50, molecular hydrogen cooling allowed primordial gas in them to collapse and form the first stars.  
Numerical simulations indicate that these stars formed in isolation (one per halo) and, due 
to the inefficiency of H$_2$ cooling, were likely very massive, from 15 - 500 $\Ms$ 
\citep{abn00, abn02, bcl99, oshea07a}.  With surface temperatures in excess of 100,000 K, many of these Population 
III (or Pop III) stars were millions of times more luminous than the sun and prodigious sources 
of both ionizing and Lyman-Werner ultraviolet (UV) radiation.  Lyman-Werner (LW) photons 
in the 11.18 - 13.6 eV energy range on average dissociate 15\% of the H$_2$ molecules they 
encounter through excitation to electronic states that decay to the vibrational continuum.  
Radiation hydrodynamical calculations indicate that the H II regions of Pop III stars were 
up to 10 kpc in diameter, with strong mass outflows capable of evicting half of the baryons 
from their halos \citep{wan04,ket04,abs06,awb07,jet07,yet07,awb07,wa07a}.  The transparency of neutral H and He to LW photons
allowed them to propagate further into the early universe than ionizing UV, dissociating H$_2$ 
far beyond the H II region.

When halos grew to greater masses they could instead cool by atomic H 
line emission.  These structures hosted the first small star populations by z $\sim$ 12 - 15. 
Although the initial mass function (IMF) of these stars remains unknown (having never been 
simulated or constrained observationally), it is thought that protogalaxies shouldered the 
bulk of cosmological reionization from z $\sim$ 8 - 15 \citep{rs01,cfw03}.  Thus, over time 
the first stars and galaxies gradually built up both an ionizing and LW dissociating UV 
background in the intergalactic medium (IGM).

A key question in cosmological reionization and structure formation is how radiation from one 
generation of stars influenced the next.  In global models of cosmological evolution that 
average over the fine structure of reionization, it has been 
suggested that the H II 
regions of the first stars and galaxies created an 'entropy floor' in which later star 
formation was discouraged \citep{oh03}.  On the other hand, numerical studies demonstrate that 
molecular hydrogen in the ionization fronts of protogalaxies \citep{rs01} enhanced gas cooling 
in the early IGM, thereby promoting star formation.  Photons from hard UV sources exhibit 
a range of mean free paths in prefront gas that broaden the ionization front.  Substantial free 
electron fractions at only a few thousand K can be established in the outer layers of these 
fronts, catalyzing H$_2$ formation on time scales much shorter than the dynamical times of the 
H II region.  These scenarios are known as negative and positive radiative feedback, respectively.

A rising LW background may have suppressed star formation in low mass halos, sterilizing them of 
molecular hydrogen needed for dynamical collapse.  \citet{met01} performed numerical simulations 
of pregalactic structure evolution in the presence of a uniform LW background and discovered that 
star formation was delayed rather than prevented.  They found that as halos grew and their central 
densities rose, their cores became self shielded from the background, permitting the formation of 
H$_2$ and their eventual cooling and collapse.  \citet{met03} later examined the effect of soft 
x-ray backgrounds due to early populations of miniquasars \citep{km05} on structure formation.  
X-ray spectra may partially ionize pregalactic clouds without raising them to high temperatures, 
creating favorable conditions for H$_2$ formation and cloud collapse.  However, this effect was 
found to be relatively mild for a wide range of x-ray backgrounds, with molecular hydrogen 
concentrations still being determined by the equilibrium between H$_2$ catalysis and Lyman-Werner
dissociation.  These numerical calculations were both performed in cosmological simulation volumes 
(1 Mpc$^3$ comoving).  \citet{oshea07b} and \citet{wa07b,wa07c} have recently revisited primordial star
formation in strong LW backgrounds in somewhat smaller cosmological volumes.

Recently, studies have turned to the suppression or promotion of star formation on small scales
in cosmological halos proximate to the first stars.  These surveys sacrifice radiation transport to follow
the collapse of the halo into a new star.  The first in this vein was \citet{oet05}, who considered 
a minihalo assumed to be 'flash ionized' (ionized on time scales much shorter than outflow times) 
by a 120 $\Ms$ Pop III star.  They found molecular hydrogen to readily form in the warm relic H II 
region, causing rapid cooling of the halo and formation of a primordial star.  The main limitation of this study was 
its neglect of the evaporation and photodissociation of the halo itself so its final state was
open to debate.  \citet{yet07} revisited local UV feedback in a broader range of environments 
and included HD cooling.  They found second generation star formation to be highly situational; when 
collapse did occur HD cooling led to fragmentation on smaller mass scales and stars significantly 
smaller than their progenitors.  More ambitious numerical attempts have been made to assemble the 
first primeval galaxies by the consecutive formation of primordial stars, one often in the relic H 
II region of its predecessor \citep{jet07,yet07}.  In these models, which emphasize the role of radiative 
feedback in protogalaxy evolution, star formation in remnant ionization fields is weakly suppressed 
and, due to HD cooling, led again to a less massive generation of stars.  

The latest experiments abandon collapse of the halo into a star and focus instead on its
dissociation and evaporation, inferring the likelihood of star formation from its final state.
\citet{su06} performed radiation smoothed particle hydrodynamical (RSPH) calculations of 
the ionization of a clump residing in the same halo as a primordial star.  Accounting for 
both photodissociation and ionization, they discovered a threshold density of 100 cm$^{-3}$ 
above which the clump could still collapse, even when the star was at 30 pc.  Central densities 
greater than this were sufficient to self shield the core of the cloud from the LW flux of the 
star.  H$_2$ forming in the I-front also protected molecular hydrogen in the cloud from 
destruction during photoevaporation.  Below these densities the radiation front generally 
overran the halo, dispersing the gas into the IGM.  In a numerical suite by \citet{s07}, halos 
with central densites above 1000 cm$^{-3}$ self shield from reasonable values of metagalactic
backgrounds as well.

The most complete survey to date of primordial halo evaporation for a grid of masses and UV fluxes 
was by \citet{as07} (hereafter AS07), who found local radiative feedback 
to be mostly neutral.  In short, halos destined to collapse into stars would still do so in 
the presence of the UV source while those too small to form stars would be prevented from 
doing so by the radiation.  They identified a novel mechanism of cloud collapse due to 
shock induced molecule formation (SIMF).  AS07 found that the D-type I-front shock sometimes 
accelerated at the center of the halo, heating it above 10,000 K and collisionally ionizing 
it.  Abrupt H$_2$ formation ensued, causing a cooling pulse leading to rapid collapse of the 
core and creation of a star.

Unfortunately, this well-parametrized study is problematic because of its choice of 
coordinate mesh.  The semianalytical halos adopted in their calculations were centered in 
a one-dimensional spherical Lagrangian grid, with incoming radiation from the outer boundary.  
In three dimensions this arrangement corresponds to a radially-symmetric halo bathed by a 
centrally directed field from all angles rather than the more realistic radiation wave that 
would sweep over an actual halo from one direction.  Two unphysical effects result from such 
implosion geometries.  First, as matter is crushed inward uniformly from all directions it 
rises to much higher densities than if compressed from one side, causing the D-type shock to 
become much stronger than in an actual photoevaporating halo.  This leads to the artificial 
heating and collisional ionization of the core evident in many of the AS07 models and 
overestimates the degree to which shocks penetrate the core.  Second, placing the core 
on the inner boundary guarantees that shocks reaching the center will rebound, in contrast to 
the one-sided compression and displacement of the core that really occurs.  Both effects lead
to serious departures from the true evolution of the halo once the shock is within 10 pc of 
the core, rendering many of the authors' findings suspect.

We present a suite of two-dimensional axisymmetric radiation hydrodynamical calculations of the ionization 
and photodissociation of cosmological minihalos by a 120 $\Ms$ star.  Our initial conditions
are spherically-averaged halos derived from cosmological initial conditions with the Enzo 
adaptive mesh refinement (AMR) code.  The baryon profiles chosen for this study are a series 
of snapshots of the same halo taken at different redshifts to evaluate the impact of the I-front on 
the halo at any of its evolutionary stages at several distances from the UV source.  The aim
of this survey is to span the range of possibilities for local radiative feedback in the 
first generation of luminous objects in the universe.  Consequently, we do not consider the
contributions of metagalactic Lyman-Werner or x-ray backgrounds here but reserve
these effects for later study.  An auxiliary objective is to evaluate the degree to which
photoevaporation of star-forming clouds is distorted in spherical geometry, a 
practice not limited to AS07 \citep{brt89,bm90,cen01}. 

In $\S$ 2 of this paper we evolve a semianalytical halo from AS07 with radiation in our two dimensional 
geometry for comparison to the one-dimensional result.  This study illustrates the sequence
of events in halo dissociation and evaporation, which is explored in $\S$ 3.  There we motivate 
our choice of halo profiles and illuminating fluxes, tabulate final outcomes for each model, 
and review the range of possiblities for radiative feedback in the cluster.  In $\S$ 4 we 
conclude.   

\section{TIS Halo Photoevaporation}

The semianalytical truncated isothermal sphere (TIS) halo model from AS07 we test here is 
2.0 $\times$ 10$^5$ $\Ms$ is located 540 pc from a 120 $\Ms$ star.  TIS halos are reasonble 
approximations of the density profile of a cosmological halo and are parametrized by their 
truncation radius r$_t$ (or outer boundary) 
\begin{align}
r_t \: = \: 102.3 \, \left(\displaystyle\frac{{\Omega}_0}{0.27}\right)^{-1/3} \, 
\left(\displaystyle\frac{h}{0.7}\right)^{-2/3} \; \notag \\
\left(\displaystyle\frac{M}{2.0 \times 10^5 \Ms}\right)^{1/3}
\left(\displaystyle\frac{1+z}{1+20}\right)^{-1} \; pc, 
\end{align}
their virial temperature 
\begin{align}
T \: = \: 593.5 \, \left(\displaystyle\frac{\mu}{1.22}\right) \; 
\left(\displaystyle\frac{{\Omega}_0}{0.27}\right)^{1/3} \, 
\left(\displaystyle\frac{h}{0.7}\right)^{2/3} \; \notag \\
\left(\displaystyle\frac{M}{2.0 \times 10^5 \Ms}\right)^{2/3}
\left(\displaystyle\frac{1+z}{1+20}\right) \; K, 
\end{align}
and their central density 
\begin{align}
\rho_0 \: = \: 4.144 \times 10^{-22} \, \left(\displaystyle\frac{{\Omega}_0}{0.27}\right) \, 
\left(\displaystyle\frac{h}{0.7}\right)^{2} \; \notag \\
\left(\displaystyle\frac{1+z}{1+20}\right)^{3} \; g \; cm^{-3}.
\end{align}
The halo density profile as a function of radius is given by 
\begin{align}
\rho (r) \: = \: \left(\displaystyle\frac{a_1}{a_2 + \xi^2} \; - \frac{b_1}{b_2 + \xi^2}\; \vspace{0.1in}
\right) \: \rho_0, 
\end{align}
where a$_1$ = 21.38, a$_2$ = 9.08, b$_1$ = 19.81, b$_2$ = 14.62, and $\xi$ = r/(r$_t$/29.4) 
from equations 103 and 104 of \citet{sir99}.

We model the evaporation of this halo with ZEUS-MP \citep{wn06, wn07a}, a massively-parallel 
Eulerian reactive flow hydrocode with self-consistent multifrequency photon-conserving UV radiative 
transfer \citep{wn07b} and the nine-species primordial gas reaction network of \citet{anet97}.  We 
have ported our algorithm to ZEUS-MP 2.0, the recent F90 public-release of the code capable of 
performing calculations in one, two, or three dimensions in cartesian (XYZ), cylindrical (ZRP), or 
spherical (RTP) coordinate meshes \citep{het06}.  Our photon-conserving transport, separate from the 
flux-limited diffusion (FLD) native to ZEUS-MP, can simulate photons from a point source centered in 
a spherical grid or plane waves along the x- or z-axes of cartesian or cylindrical boxes.  Depending
on the dimensionality and type of coordinate mesh, the code can apply spherical shell, conjugate
gradient (CG), or multigrid methods to solve Poisson's equation for the self-gravity of the gas. 

We centered the TIS halo at the origin of a two-dimensional axisymmetric cylindrical (ZR) coordinate 
box, with boundaries of -125 pc and 125 pc in z and 0.01 pc and 125 pc in r.  The grid was discretized 
into 1000 zones in z and 500 zones in r for a spatial resolution of 0.25 pc.  Outflow conditions were 
applied to the upper and lower z boundaries and reflecting and outflow conditions were assigned to the 
inner and outer boundaries in r, respectively.  The gas was primordial, 76\% H and 24\% He by mass.  
The gravitational potential of the dark matter was included by computing the potential necessary to 
cancel pressure forces everywhere on the grid (setting the halo in hydrostatic equilibrium) and holding
this potential fixed throughout the simulation.  Excluding dark matter dynamics introduces no significant 
errors because the gas in the halo evolves on much shorter time scales than either the Hubble time or 
merger time scales.  Updates to the self gravity of the gas were performed every hydrodynamical time step 
by solving Poisson's equation with a two dimensional CG solver.  Cooling by electron collisional excitation 
and ionization, recombination, bremsstrahlung, and inverse Compton scattering of the cosmic microwave 
background (CMB), assuming a redshift $z = $ 20, was present in all the models.  H$_2$ cooling was also 
included using the cooling curves of \citet{gp98}.  

We partition photon 
emission rates by energy into 40 uniform bins from 0.755 eV to 13.6 eV and 80 logarithmically spaced 
bins between 13.6 eV and 90 eV according to the blackbody spectrum of the 120 $\Ms$ star.  The rates 
are normalized by the total number of ionizing photons emitted per second by the star from \citet{s02}.  
Photons in the lower energy range cannot ionize H or He but do drive a host of chemical reactions that 
regulate H- and H$_2$ formation and are tabulated in Table 1 of \citet{wn07b}.  H$_2$ photodissociation 
rates are computed with the self-shielding functions of \citet{db96} modified for thermal doppler
broadening as in AS07 to account for gas flows.  The legitimacy of thermal broadening as a proxy for  
flows in evaporating halos is uncertain, but including it reduces shielding at intermediate column 
densities, setting upper limits on H$_2$ dissociation.  We attenuate the intensity of the plane wave 
by 1/R$^2$ to approximate geometrical dilution of the radiation.  The cloud is irradiated for 2.5 Myr,
the main sequence lifetime of the 120 M$_\odot$ star, and then left to evolve in the relic H II region
for another 2.5 Myr.

\begin{figure*}
\epsscale{1.15}
\plotone{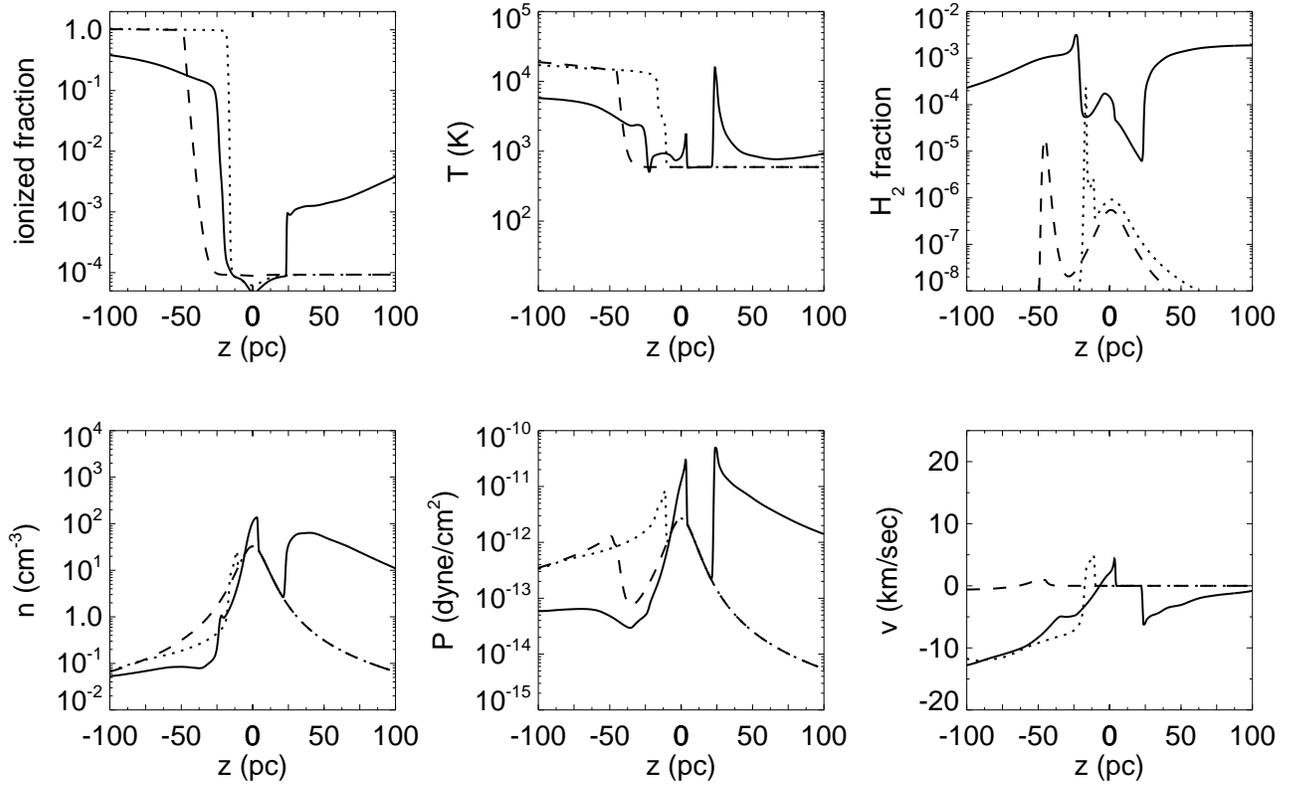}\vspace{0.15in}
\caption{Ionized fraction, density, temperature, pressure, H$_2$ fraction and velocity profiles in
the TIS halo.  Dashed line: 200 kyr, dotted line: 2.4 Myr, solid line: 4.5 Myr. \label{fig:ahn1}} 
\vspace{0.075in}
\end{figure*}

As a control we chemothermally evolved the halo for 4.8 Myr without radiation to determine the 
degree to which the core cools and condenses in the absence of stellar feedback.  Although e- and 
H$_2$ fractions are initially 
uniform throughout the cloud, H$_2$ catalysis and cooling follows the density profile and is 
greatest in the core where the density peaks.  Cooling disrupts the pressure balance with
gravity and the core begins to contract.  As densities rise so do molecular hydrogen production 
and cooling, and the collapse accelerates.  In this halo infall is leisurely, and we find that 
central densities only change from 31 cm$^{-3}$ to 36 cm$^{-3}$ in 4.8 Myr, consistent with what
was found in AS07.  The self gravity of the gas is essential to this process; if deactivated, little gas 
motion occurs on these time scales.  Ahn \& Shapiro find that if collapse proceeds undisturbed 
a star forms in this halo at 32 Myr, so it is an appropriate candidate for radiative feedback.

\subsection{The R-Type front}

\begin{figure*}
\epsscale{1.15}
\plotone{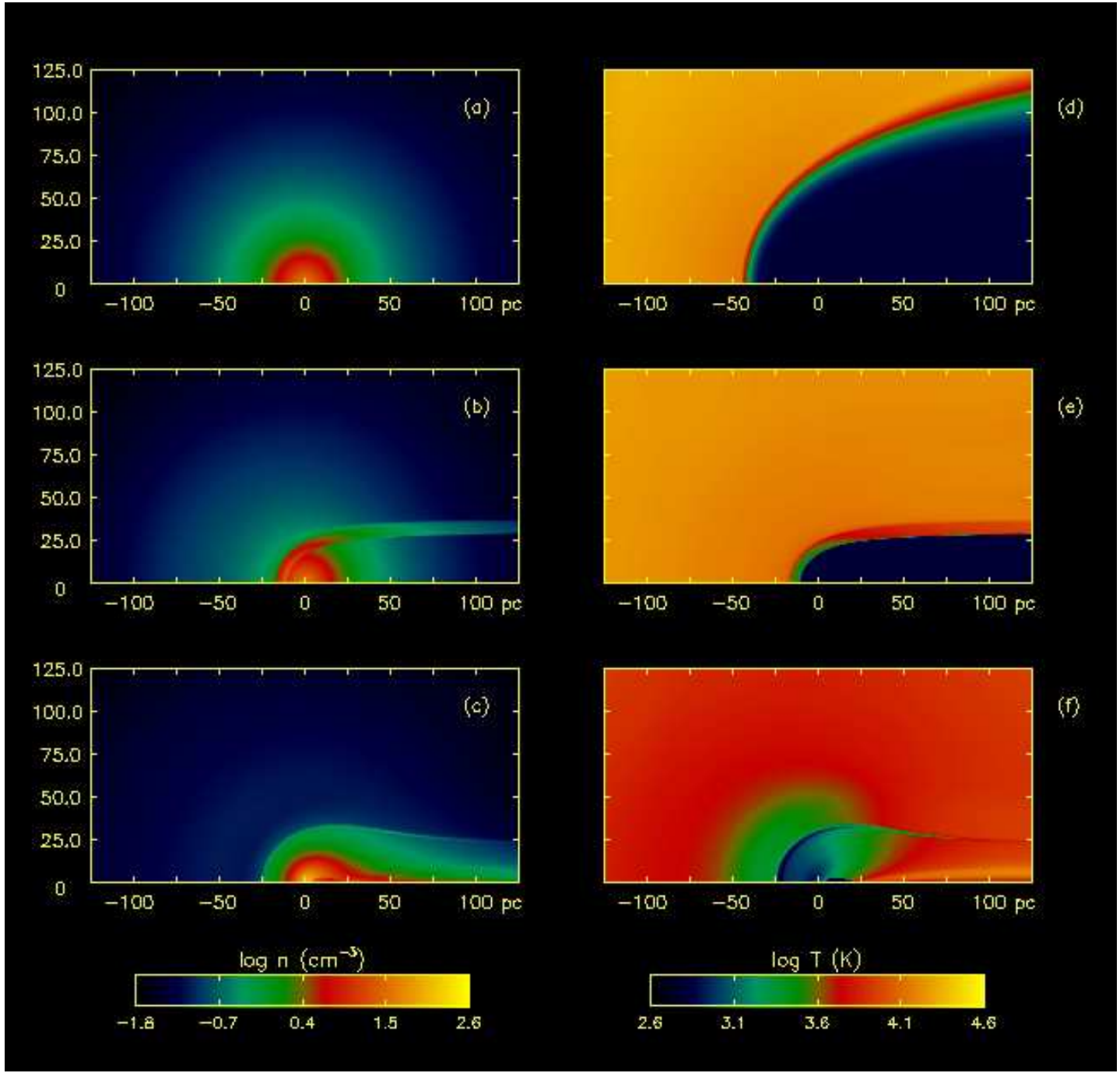}\vspace{0.15in}
\caption{The photoevaporating TIS halo.  Panels (a), (b), and (c) are densities at 200 kyr, 2.4 Myr,
and 4.5 Myr, respectively.  Panels (d), (e), and (f) are temperatures at 200 kyr, 2.4 Myr, and 4.5 
Myr, respectively.  In this and all other color figures in the paper, the horizontal axis is $Z$ and 
the vertical axis is $R$. \label{fig:ahn2}} 
\vspace{0.075in}
\end{figure*}

We show ionized fraction, number density, temperature, pressure, H$_2$ fraction, and velocity 
profiles along the $z$-axis through the center of the TIS halo at three times in Fig \ref{fig:ahn1}.  
Figs \ref{fig:ahn2} and \ref{fig:ahn3} are images of the densities, temperatures, and H$_2$ fractions on the 
grid for the same three times. The I-front enters the lower $z$ face of the box at t = 0.  The 
absence of shock plateaus in the abrupt temperature and pressure profiles together with the 
undisturbed densities at 200 kyr indicate that the front is R-type. Lyman-Werner photons ahead of 
the front dissociate molecular hydrogen both in the core and beyond, as seen in the sudden 
displacement downward of the H$_2$ profile from 2.0 $\times$ 10$^{-6}$ by t =200 kyr.  There are
two peaks in this curve, one at -50 pc and one at the core.  The first (2 $\times$ 10$^{-5}$) is 
H$_2$ in the outer layer of the I-front, whose partial ionization and moderate temperatures  
facilitate its formation.  As discussed below, this H$_2$ layer eventually acts as a 
shutter, preventing LW flux from the star from reaching the core.

The second peak is the equilibrium H$_2$ abundance ($\sim$ 6 $\times$ 10$^{-7}$) set in the core 
by a balance between  
between Lyman-Werner dissociation and formation through the H- channel.  Equilibrium fractions 
fall beyond the core because the large central densities replenish H$_2$ at greater rates than 
in the more tenuous outer envelope.  Molecular hydrogen fractions ahead of the front remain 
stable while it is R-type but begin to rise when it becomes D-type at $\simeq$ 30 pc.  The front 
is broad in the diffuse IGM densities 50 pc from the center of the halo ($\sim$ 25 pc at 200 kyr) 
due to the spread of mean free paths in the hard UV tail of the 100,000 K blackbody spectrum.  
It narrows and slows as it ascends the density gradient toward the core of the cloud. 

Panels (d) of Fig \ref{fig:ahn2} and (a) of Fig \ref{fig:ahn3} reveal the cometary appearance of 
the front at 200 kyr due to its preferential advance in the stratified layers above and below the 
halo. This parabolic shadow, common to all our models, is predicted analytically using
inverse Str\"{o}mgren layer arguments (section 2.3.2 of \citet{sir04}).  Molecular 
hydrogen in the R-type front extends in a yellow arc from the $z$-axis out to the end of the box on
the right in panel (a) of Fig \ref{fig:ahn3}.  The reddish hues of the higher equilibrium H$_2$ 
fractions within the halo stand out against the greenish hues of the lower fractions in the diffuse 
envelope.  The densities remain largely unchanged because the front is still supersonic.

\subsection{The D-Type Front}

As the front slows, a pressure wave builds in the ionized gas, pushing past the front and
steepening into a shock.  This marks the transformation of the front to D-type and the onset 
of photoevaporation.  The shock soon detaches from the front as shown in the t = 2.4 Myr 
temperature plot in Fig \ref{fig:ahn1}.  The gas drops sharply in temperature from 20,000 K 
to 3000 K from ionized to shocked gas.  The neutral shell extends a few parsecs to the shock 
where the temperature again drops, this time to the background.  The shocked shell is the sharp 
peak to the left of z = 0 in the density profile at 2.4 Myr.  As the front approaches the center 
of the cloud ionized backflow to the left is visible in the velocity profile at 2.4 Myr.  This 
photoevaporated gas exits at 5 - 10 km/sec, well above the 2 - 3 km/sec escape velocity of the 
halo, and does not return unless recaptured by mergers, whose time scales at $z \sim$ 20 are 
$\sim$ 20 Myr.  The shock is 13 pc from the core at a velocity of 5 km/sec when the star exits 
the main sequence.

\begin{figure}
\resizebox{3.45in}{!}{\includegraphics{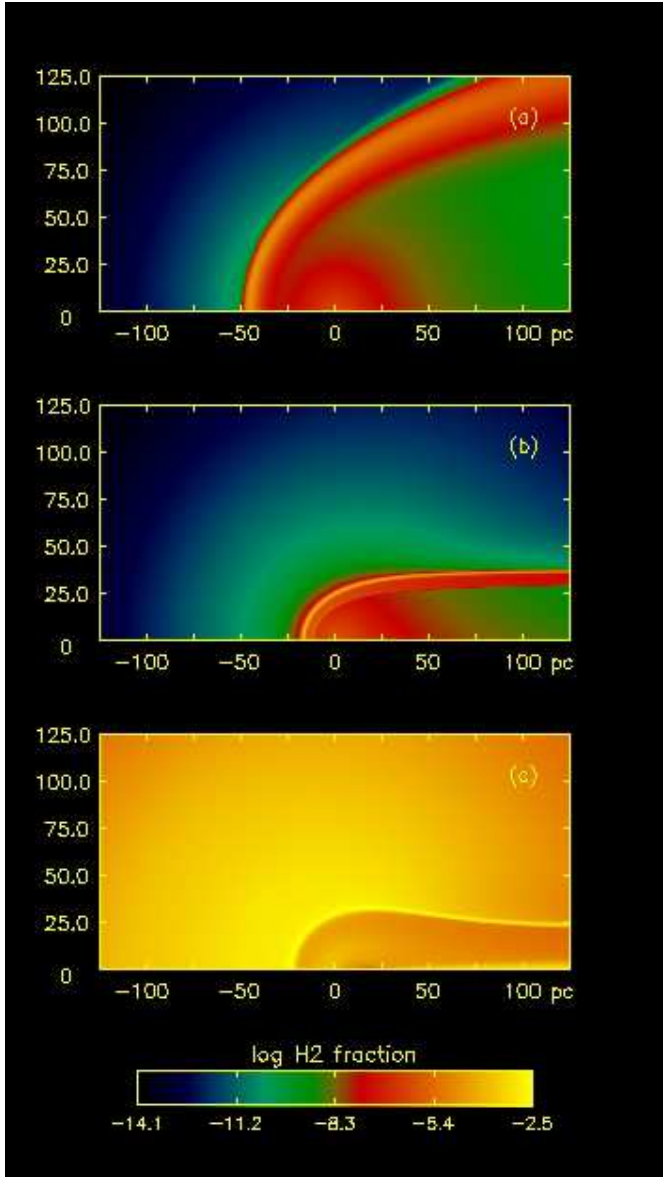}}
\caption{H$_2$ fractions in the photoionized TIS halo: panels (a), (b) and (c) are at 200 kyr, 
2.4 Myr, and 4.5 Myr, respectively.} \vspace{0.25in}
\label{fig:ahn3}
\end{figure}

The high energy photons in the front encounter greater densities as they penetrate the 
neutral shocked shell.  The electrons they liberate catalyze molecular hydrogen formation at 
the base of the shell, elevating H$_2$ fractions there to 3.0 $\times$ 10$^{-4}$.  As the  
D-type front advances from -30 pc to -25 pc molecular hydrogen fractions in the halo steadily 
rise and then level off as they come to new equilibrium values, being partially shielded from 
the LW flux.  Core H$_2$ fractions reach 1.0 $\times$ 10$^{-6}$ and remain there until the death 
of the star.  As molecular hydrogen levels rise, temperatures in the core fall from 600 K to 400 
K.  This modest restoration of H$_2$ falls short of the original fraction that was dissociated.  
The initial free electron fraction on the grid (which drives H$_2$ formation at the center) is 
slowly depleted by recombinations as the halo evolves.  Recombinations are faster in the higher 
densities of the core, producing the dip in ionized fraction there at 2.4 Myr.  Note that in the 
halo beyond the front the H$_2$ fraction generally follows the density profile, as one would 
expect.

In panel (e) of Fig \ref{fig:ahn2} the partially evaporated halo casts a nearly cylindrical 
shadow along the $z$-axis by the time the Pop III star dies.  Ionized gas surrounding the shadow 
drives an axial implosion shock toward the axis.  The ablation shock approaching the halo is 
the green parabola that begins on the axis and narrows outward along the arc in the temperature 
image.  H$_2$ cooling is visible in the core as the circular blue patch at the center of the mesh; 
corresponding elevated molecular hydrogen concentrations are visible in yellow in panel (b) of 
Fig \ref{fig:ahn3}.  Densities are greatest in the shock and the center of the cloud in panel (b) 
of Fig \ref{fig:ahn2}.  The cometary shocked shell being driven toward the axis is the thin layer 
of green and aqua parallel to the $z$-axis.  At 2.4 Myr the left hemisphere of the cloud has been 
largely photoevaporated and the inner part of the halo is compressed to the right. 

\subsection{The Relic H II Region}

The H II region begins to recombine when the 120 $\Ms$ star dies at 2.5 Myr, most likely by direct 
collapse to a black hole \citep{het03}.  H$_2$ formation becomes explosive in the warm 
relic ionization field and shock, which continues into the core as it slows.  As ionized 
gas surrounding the shadow of the halo cools some pressure support for the shock is lost but 
it still converges on the $z$-axis.  After the death of the star a rarefaction wave develops in 
the opposite direction just behind the shock and retreats to -37 pc by 4.5 Myr, as seen in the
dip in pressure there in Fig \ref{fig:ahn1}.

At 4.8 Myr the shock, highly enriched with molecular hydrogen, merges with the core.  The 
central density rises to 398 cm$^{-3}$, more than ten times its original value, with an H$_2$ 
fraction of 3.0 $\times$ 10$^{-3}$.  In the absence of any radiation this core would have 
chemothermally evolved to a central density of only 36 cm$^{-3}$.  While both density and 
H$_2$ have been greatly enhanced by the 
radiation, the gas is moving at 5 km/sec.  Although it will likely slow below 2 - 3 km/sec
and become gravitationally bound to the halo, it is unclear if a star will form in this gas.

The cylindrical shock converges symmetrically on the $z$-axis on all azimuths, heating to over
10,000 K and becoming collisionally ionized to the right of the coordinate origin in the 4.5 
Myr temperature and ionized fraction profiles.  A surge in molecular hydrogen production 
follows along the central axis, visible in both the H$_2$ fraction plot for z $>$ 25 pc and 
panel (c) of Fig \ref{fig:ahn3}.  The shock converges on the axis first in the lower densities
beyond the core; as gas closer to z = 0 reaches the axis a reverse shock builds toward the halo 
in the negative z-direction as shown in the velocity profile at 4.5 Myr.  This shock drives 
material back into the halo but is still 25 pc from the core at 4.8 Myr.  The pulse of H$_2$ 
formation is the SIMF found by Ahn \& Shapiro in many of their models.  Actual halos have 
morphologies that would break simultaneous convergence of gas onto the axis and probably avoid 
SIMF.  

When the H II region begins to recombine, the ionized backflow to the left of the halo is at 
first isothermal, with a density gradient that falls with distance.  The isothermal equation of
state mandates pressure gradients wherever there are density drops, and these pressures 
accelerate the backflow even as the gas recombines, as shown in the 4.5 Myr velocities 
for z $<$ 0.  Also, the lowest temperatures at later times are just behind the shock remnant 
merging with the core from the left; these correspond to the pressure minimum and are due to 
PdV work done by the rarefaction wave receding from the core.  In general, temperatures and 
ionization fractions drop fastest where densities are greatest, as in these plots.

The density and temperature images in panels (c) and (f) reveal the slight displacement of 
the core to the right of the coordinate center by the remnant shock at 4.5 Myr.  A 
reflection shock from the axis beyond z = 25 pc collides with partially ionized gas still 
driven inward.  The warm relic H II region encircles the eroded halo in teardrop density 
contours and the shock, now cooled to 160 K, appears as a single large tail curled around 
the cool dense halo core.  Molecular hydrogen production radiates outward from the core in 
a yellow arc that dims slightly with distance from the center of the mesh in panel (c) of 
Fig \ref{fig:ahn3}.  This is due to the density dependence of H$_2$ formation rates:  
production erupts first in the core, following later in the more diffuse outer envelope.

\begin{figure*}
\epsscale{1.15}
\plottwo{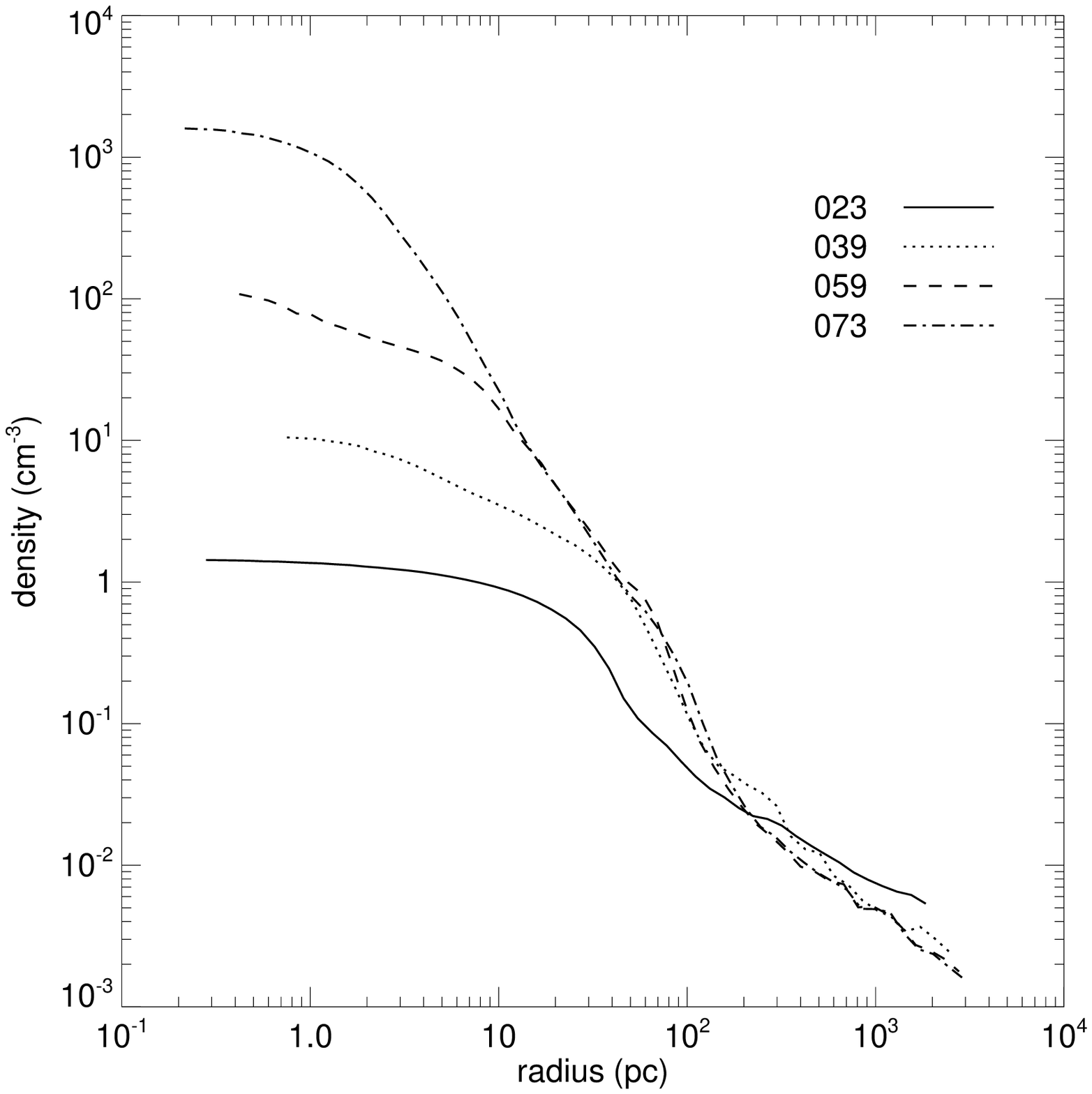}{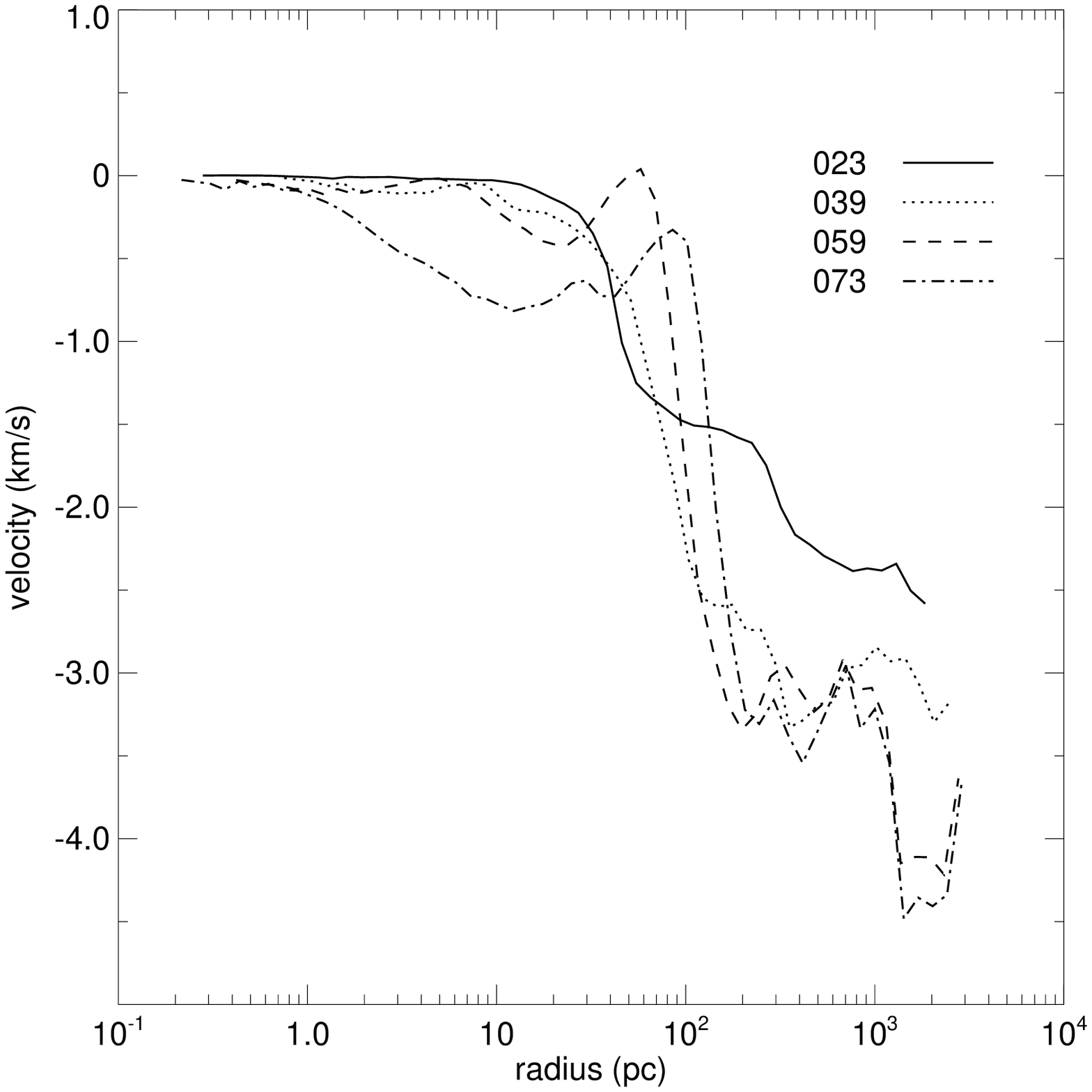}\vspace{0.15in}
\caption{The four halo profiles.  The redshifts of the 023, 039, 059, and 073 profiles are 23.9, 
17.7, 15.6, and 15.0, respectively.  Left: densities.  Right: radial velocities. \label{fig:halop}} 
\vspace{0.075in}
\end{figure*}

We note that a more realistic point source of radiation situated on the $z$-axis would cast a 
parabolic shadow like the one at later times in the \citet{su06} calculations rather than the 
cylindrical shadow in our simulations.  However, recombination photons, excluded in our models
by the on-the-spot approximation, may compensate by photoionizing gas toward the central axis.  
A shock would still be driven into the shadow, narrowing it to a greater degree than in point 
source simulations that ignore reprocessed radiation but perhaps not to the extent in UV plane 
wave geometries like ours.  The shadow becomes increasingly cylindrical with point sources at 
greater distances.

\subsection{Consequences of One-Dimensional Geometries}

Figs 1 of \citet{wn07b} and 8 of AS07 illustrate the general agreement that would be 
expected between the two models while the front is far from the core, supersonic and still 
free of curvature effects.  The fraction profiles for all nine species indicate that both 
codes predict the same position for the front at t = 0.6 t$_\ast$, where t$_\ast$ is the 
main sequence lifetime of the star.  Furthermore, in both studies the I-front becomes fully 
D-type $\sim$ 25 pc from the center of the halo and is 10 - 12 pc from the core when the 
star dies.

As matter closes in on the core from all sides in the AS07 code the two solutions begin to
diverge.  When the shock in the one-dimensional code reaches the density plateau near the
center it actually accelerates, which would not occur in a core compressed from one side.  
In our two-dimensional simulations the shock gradually decelerates from 7.5 km/sec to 5 
km/sec as it penetrates the core.  Disagreement between the two models becomes acute as the shock 
artificially heats and collisionally dissociates the core in the Lagrangian code.  A 
cascade of thermal events at the center of the cloud follows, ending in spurious runaway 
collapse.  Rapid H$_2$ formation follows collisional ionization, in turn driving excessive 
cooling that leads to isothermal jump conditions in the shock remnant.  The density jump in 
the remnant rises to very high values that further accelerate cooling and collapse.  

The velocity of the shock in our model tapers as the rarefaction wave drains 
it of pressure support and 
it snowplows gas in the core.  Densities rise when the remnant reaches the core but not 
to the degree associated with the artificial cooling and compression in the one-dimensional 
calculation.  H$_2$ fractions rise at the center because of mixing with molecular hydrogen 
accumulated in the remnant and formation by \textit{in situ} H- rather than collisional ionization 
by an enhanced shock.  Radiative
feedback in the end result is unclear, unlike the severe positive feedback predicted by AS07.

\begin{figure*}
\epsscale{1.15}
\plotone{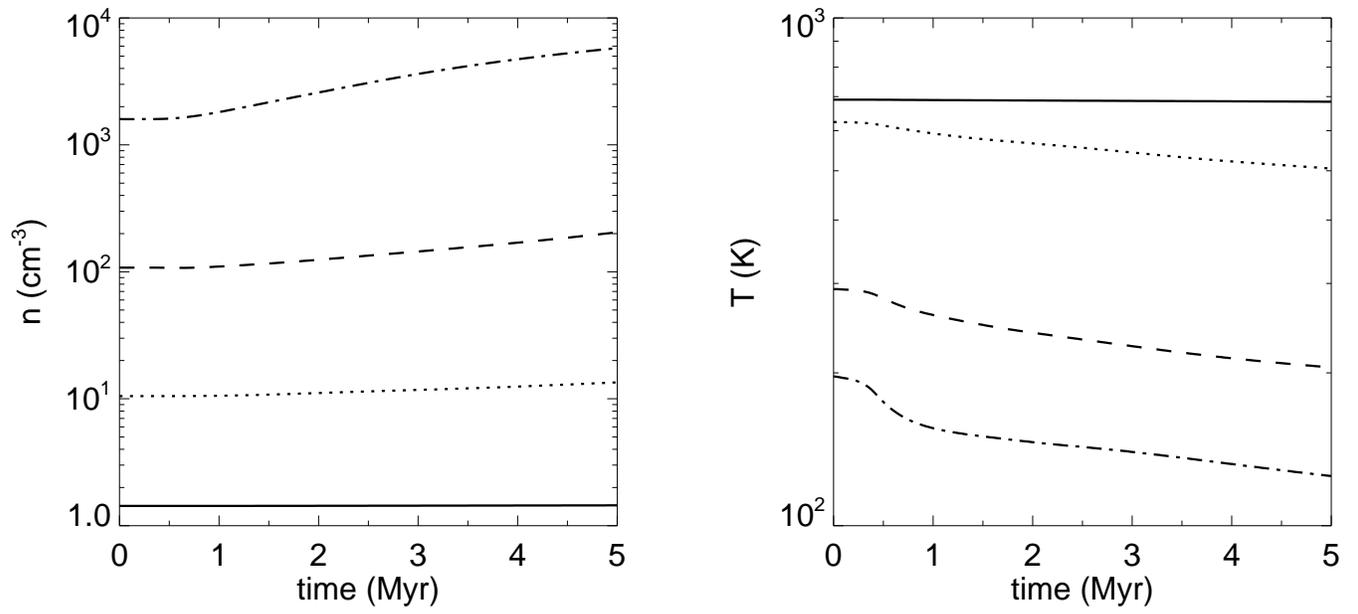}\vspace{0.15in}
\caption{Central densities and temperatures over 5 Myr in the absence of external radiation.  Solid:
023 halo; dotted: 039 halo; dashed: 059 halo; dot-dashed: 073 halo. 
\label{fig:norad}} 
\vspace{0.075in}
\end{figure*}

Contrary to the claim of AS07, the bounce of the shock from the inner reflecting boundary in 
their model cannot be equated to a planar shock interacting with the central density peak.  
The remnant instead displaces the gas at the center without the rebound associated with core
bounce.  Our model thus avoids the unphysical delays in core collapse that follow from the 
reflection of the shock at the center of the AS07 models.

We emphasize that these difficulties are not unique to AS07 but are general to one dimensional
spherically-symmetric models of the external photoevaporation of star forming clouds \citep{brt89,bm90,cen01}.

\section{Two-Dimensional Enzo Models}

The initial conditions used in our two-dimensional calculations are taken from a simulation 
performed with Enzo, a publicly available, extensively tested adaptive mesh 
refinement cosmology simulation code \citep{enzomethod, enzocomp}.  The details of this 
calculation (and others in the series) are described in detail in \citet{oshea07a} but are 
summarized here for clarity.  We select one calculation from the suite of 12 in O'Shea \& 
Norman 2007, choosing simulation L0\_30D since it has the smallest halo mass at the epoch of
\begin{figure*}
\epsscale{1.15}
\plotone{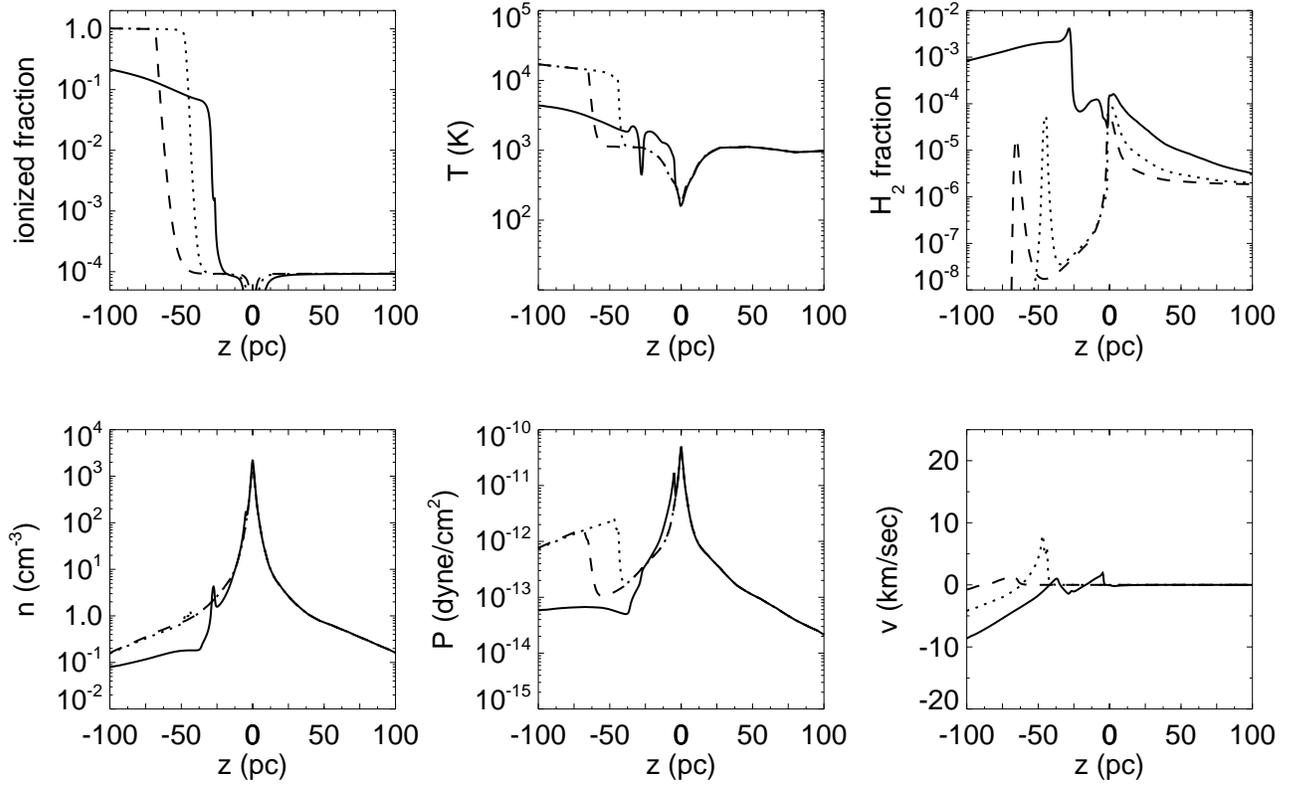}\vspace{0.15in}
\caption{Ionized fraction, density, temperature, pressure, H$_2$ fraction and velocity profiles for
the 073\_500pc model. Dashed line: 200 kyr (the R-type front), dotted line: 800 kyr (the D-type front), 
solid line: 5.0 Myr (the relic H II region). \label{fig:073500pcpr}} 
\vspace{0.075in}
\end{figure*}
\begin{figure*}
\epsscale{1.15}
\plotone{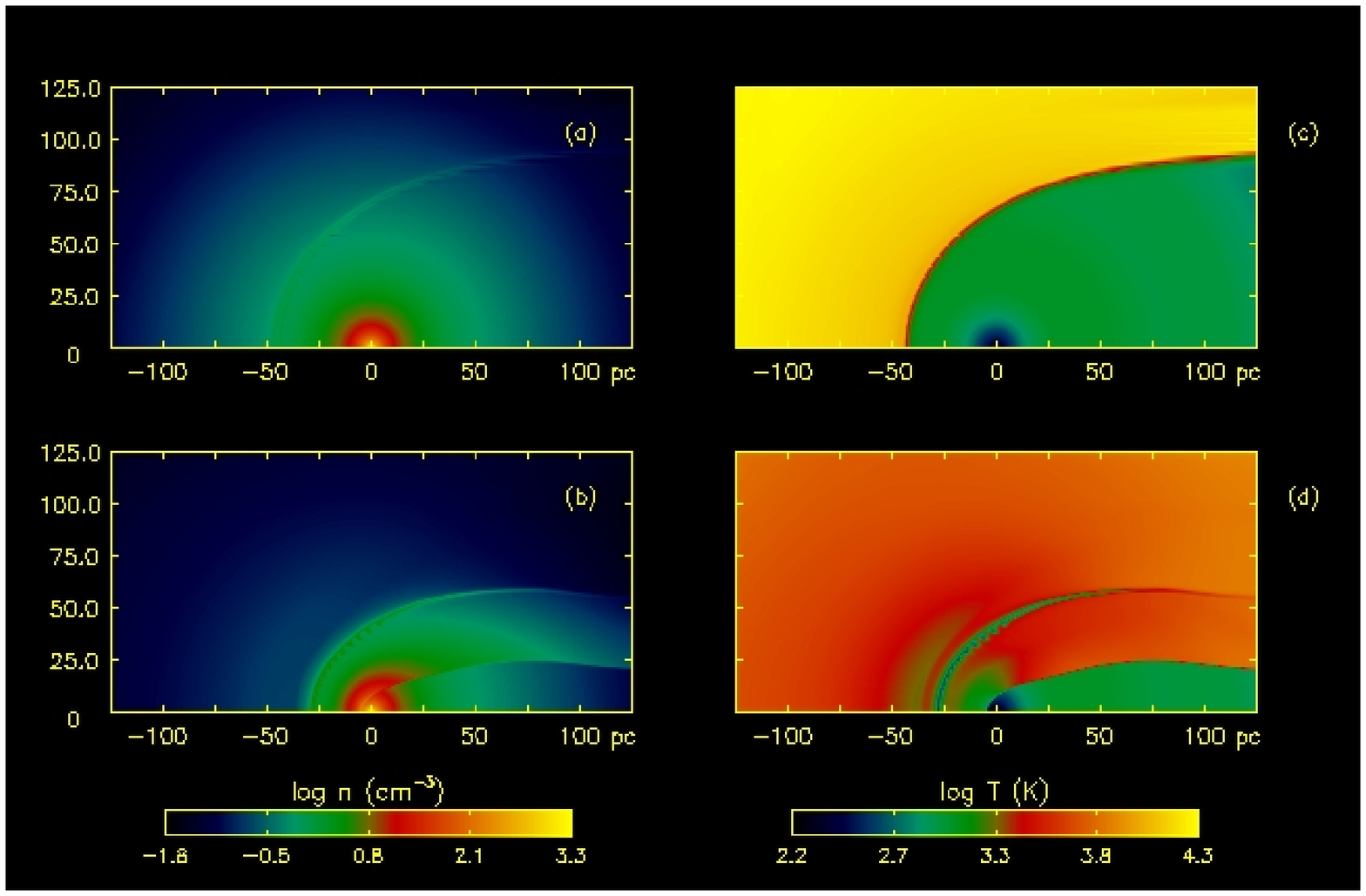}\vspace{0.15in}
\caption{Halo evaporation: model 073\_500pc.  Panels (a) and (b) are densities at 800 kyr and 
5.0 Myr, respectively.  Panels (c) and (d) are temperatures at 800 kyr and 5.0 Myr, respectively. 
\label{fig:073500pcrhoT}} 
\vspace{0.075in}
\end{figure*}
collapse ($1.35 \times 10^5$~M$_\odot$). Here, the epoch of collapse is defined to be the redshift
in the simulation at which gas in the given halo has collapsed to central densities for which star 
formation is inevitable.  This simulation was initialized at z=99 in a 
cosmological volume 300 h$^{-1}$ kpc (comoving) on a side, with a $128^3$ root grid and 3 
static nested grids, and an effective resolution of $1024^3$ cells/particles on the highest
level static grid.  This corresponds to a dark matter (gas) resolution of $2.60 (0.40)
$~M$_\odot$.  The highest level static nested grid is large enough to contain the entire  
Lagrangian volume which will end up within the virial radius of the final halo, ensuring that 
the halo is resolved by approximately $5 \times 10^4$ dark matter particles and a comparable 
number of cells for the equations of hydrodynamics.  

\begin{figure}
\resizebox{3.45in}{!}{\includegraphics{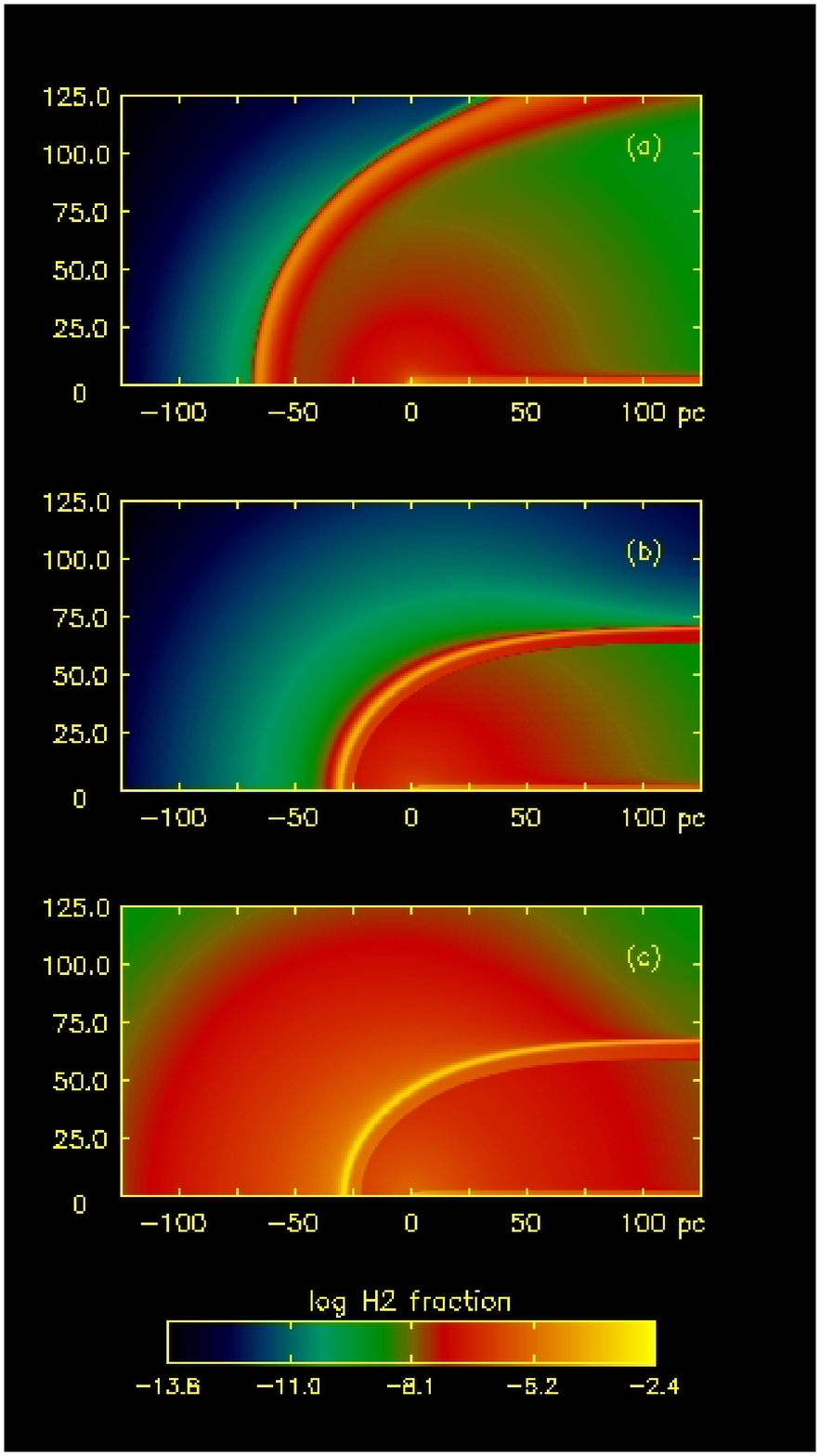}}
\caption{073\_500pc model: H$_2$ fractions.  Panels (a), (b) and (c) are at 210 kyr, 
2.29 Myr, and 2.59 Myr, respectively.} \vspace{0.25in}
\label{fig:073500pcH2}
\end{figure}

\begin{figure*}
\epsscale{1.15}
\plotone{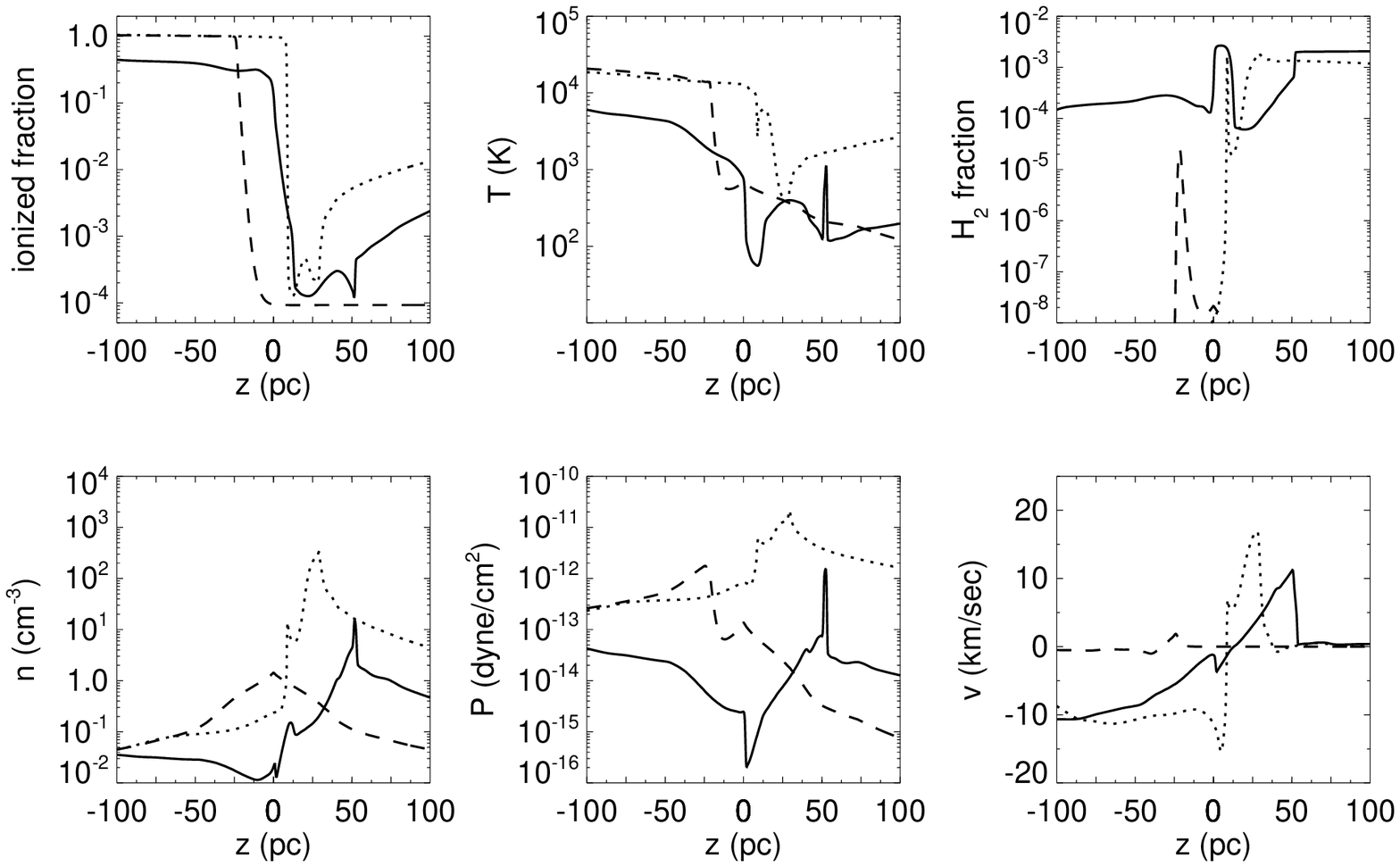}\vspace{0.15in}
\caption{Ionized fraction, density, temperature, pressure, H$_2$ fraction and velocity profiles for
the 023\_500pc model. Dashed line: 225 kyr (the R-type front), dotted line: 2.5 Myr (the D-type front), 
solid line: 5.0 Myr (the relic H II region). \label{fig:023500pcpr}} 
\vspace{0.075in}
\end{figure*}

\begin{figure*}
\epsscale{1.0}
\plotone{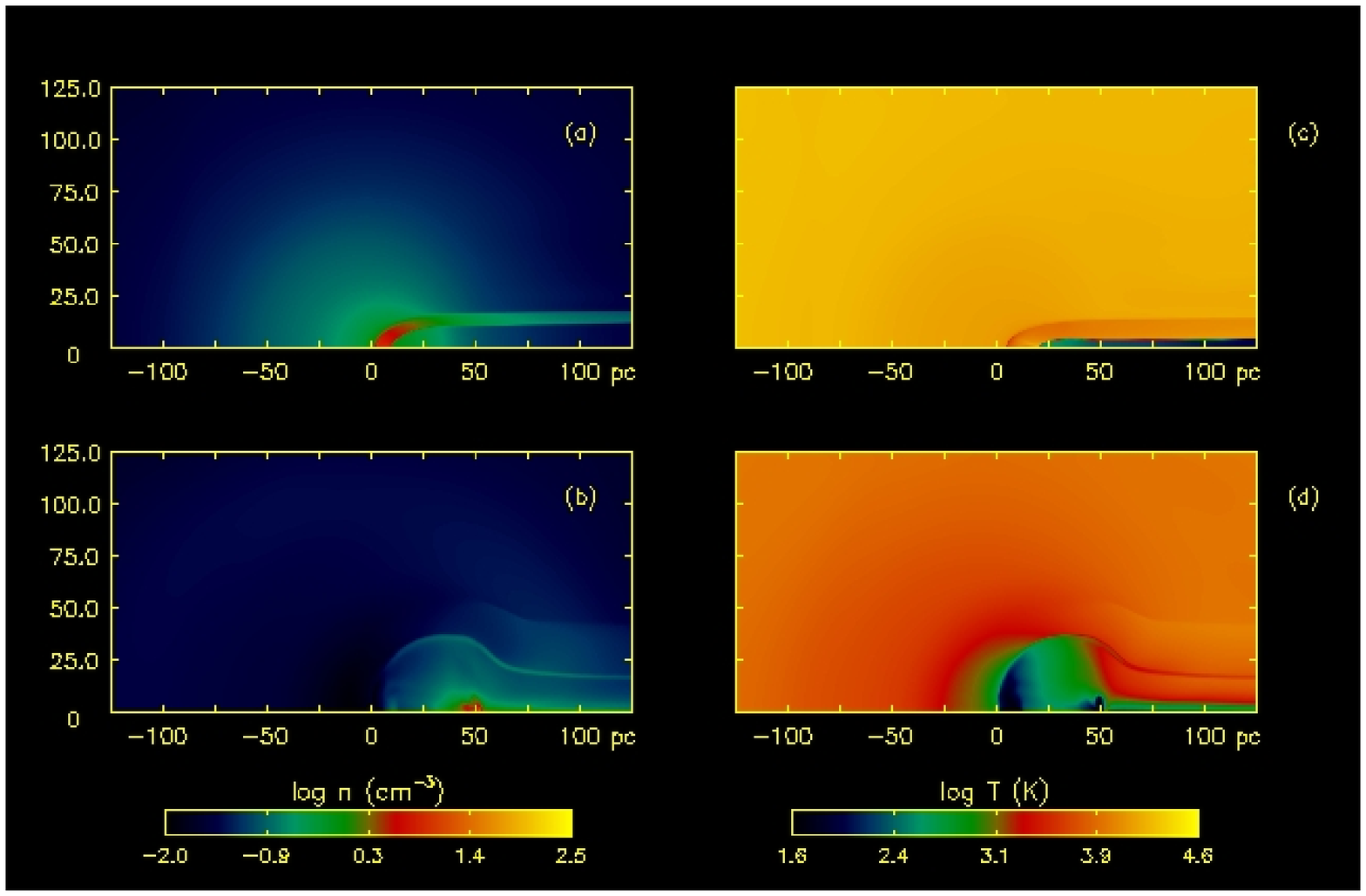}\vspace{0.15in}
\caption{Halo evaporation: model 023\_500pc.  Panels (a) and (b) are densities at 1.69 Myr and 
5.0 Myr, respectively.  Panels (c) and (d) are temperatures at 2.02 Myr and 5.0 Myr, respectively. 
\label{fig:023500pcrhoT}} 
\vspace{0.075in}
\end{figure*}

This calculation is then evolved from $z=99$ until the collapse of the most massive halo in 
the simulation following the equations of dark matter dynamics, hydrodynamics, and a 9-species 
nonequilibrium primordial chemistry model \citep{abet97,anet97}.  We employ a maximum of 28 
levels of adaptive mesh refinement, until the gas at the center of the halo cools 
and collapses to high (n$_H > 10^8$~cm$^{-3}$) densities at $z = 24.74$.  Cells are refined 
on a variety of criteria, including dark matter and baryon density, cooling time, Jeans length, 
and shock and energy gradient resolution.  In addition, mass is refined in a super-Lagrangian  
way, such that the gas mass resolution at the maximum level of refinement is $m_{cell} \simeq 
10^{-3}$~M$_\odot$.  Data is output periodically throughout the simulation, and 
then more rapidly during the collapse of the gas in the halo, such that one 
output is generated every time the central halo density increases by a factor of 8.  
One-dimensional profiles are then created by finding the cell with the highest baryon density 
in the collapsed halo core and taking mass-weighted spherical averages of baryon density, 
temperature, and other quantities.  We use logarithmic bins so that the spatial extent of the 
bins in the radial profile always approximates the cell resolution at any given radius.  Four
of these spherical profiles are then imported into ZEUS-MP.

\begin{deluxetable}{ccccc}
\tabletypesize{\scriptsize}
\tablecaption{ Halo Photoevaporation Models\label{tbl-1}}
\tablehead{
\colhead{n$_c$} & \colhead{150 pc} & \colhead{250 pc} & \colhead{500 pc} & \colhead{1000 pc}}
\startdata
 1.43 cm$^{-3}$ & 023\_150pc & 023\_250pc & 023\_500pc & 023\_1000pc \\
 10.5 cm$^{-3}$ & 039\_150pc & 039\_250pc & 039\_500pc & 039\_1000pc \\
 108  cm$^{-3}$ & 059\_150pc & 059\_250pc & 059\_500pc & 059\_1000pc \\
 1596 cm$^{-3}$ & 073\_150pc & 073\_250pc & 073\_500pc & 073\_1000pc \\
\enddata
\end{deluxetable}

The four evolutionary stages of the 1.35 $\times$ 10$^5$ $\Ms$ primordial halo used in our study 
are shown in Fig \ref{fig:halop}.  Their central densities range from 1.43 cm$^{-3}$ to 1596 
cm$^{-3}$.  We consider consecutive profiles from a single halo rather than sampling the entire 
cluster at a fixed redshift for three reasons.  First, the halos in the cluster have similar 
profiles and are largely coeval, so the emergent I-front may encounter a narrower range of 
density gradients at a single redshift than in one halo over a range of redshifts.  Second, the 
time at which the cluster is engulfed by the expanding H II region is an open parameter. Finally,
we avoid constraining our results to a single cluster of halos and its associated properties 
(number of peaks, radius, etc).  The possibilities for radiative feedback in the entire cluster 
are better explored by evaporating a single halo whose central densities vary from low values 
that are easily ionized to high values in which core collapse would proceed uninterrupted. 

This halo mass was selected because it is the smallest in which a star would be expected to form.  
The ionization front would have less impact on more massive halos, so feedback effects would be 
most prominent in this one.  This is roughly twice the minimum halo mass chosen by AS07, who adopted 
the smallest halo capable of collapsing to core densities of 10$^8$ cm$^{-3}$ (which marks the 
onset of three-body H$_2$ production and runaway cooling) in a Hubble time.  A more self-consistent 
choice would be the minimum halo mass that could reach these densities before being disrupted by 
mergers, $\sim$ 20 Myr at z $\sim$ 20.  

Each profile was illuminated by a 120 $\Ms$ star at four distances: 150, 250, 500, and 1000 pc.  
This star's blackbody spectrum is representative of those in the 100 - 500 $\Ms$ mass range.
These distances exceed the actual range of separations between star and satellite in the chosen 
Enzo simulation (200 pc - 500 pc) but are typical of the other clusters formed in the \citet{oshea07a} 
survey.  Feedback from lower mass Pop III stars (30 - 80 $\Ms$) warrant separate study because their 
spectra are not as hard as those of very massive stars and may drive different chemistry in the halos.  
They also illuminate them for longer times (but not for merger times that would require dark matter 
dynamics in the simulations).  We tabulate our grid of models in Table \ref{tbl-1}.

We employ the same computational box, resolution, and boundary conditions used  
in $\S$ 2.  The halos are again set in hydrostatic equilibrium by computing the gravitational 
potential necessary to cancel pressure forces in the gas.  As before, this potential is held 
constant over the entire simulation while the self-gravity of the gas is evolved throughout. 
The halos do exhibit infall velocities as shown in Fig \ref{fig:halop} but they are 
relatively mild.  Their inclusion would only slightly aid positive feedback effects and be 
irrelevant to negative feedback or the destruction of the halo.  

For simplicity, we assume ionization and H$_2$ fractions of 1.0 $\times$ 10$^{-4}$ and 2 
$\times$ 10$^{-6}$, respectively, appropriate for the IGM at z $\sim$ 20.  Free 
electron fractions can be depressed below this value in the cores of denser halos by 
recombinations.  This can lead to minor errors in equilibrium H$_2$ fractions if the halo 
is photodissociated because they are set by the balance between destruction by LW flux and 
production by the H- channel.  Uniform ionization fractions can lead to overestimates of H- 
in the core, but H$_2$ fractions are so low when the halo is dissociated that cooling is 
effectively halted there anyway.  Recombinations in the core soon reduce free electron 
fractions to levels approximating those in the Enzo models, mitigating initial overestimates.

We somewhat underestimate cooling in the core in the course of photoevaporation by imposing
uniform H$_2$ fractions but this too results only in minor inaccuracies.  Diffuse halos are 
dissociated by Lyman-Werner flux (and would condense very little in 5 Myr  
without the star).  The free electron fractions very rapidly establish
H$_2$ concentrations approximating those in the Enzo halos in denser halos shielded from LW
photons, so cooling proceeds nearly uninterrupted.  Initial H$_2$ fractions in halos of 
intermediate densities that are only partially dissociated are reset to the equilibrium 
established by production through the H- reaction and destruction by LW photons.  H$_2$ 
cooling is not very prominent in such halos and is halted if core H$_2$ fractions fall 
below 1 $\times$ 10$^{-4}$.

In Fig \ref{fig:norad} we show the evolution of central densities and temperatures in all four
halo profiles over 5 Myr including Enzo H$_2$ fractions, the uniform ionization fraction and no 
radiation.  H$_2$ cooling causes little changes in the 023 and 039 halos because cooling rates 
are small in their diffuse cores.  Both free-fall and cooling times are much shorter in the 
denser profiles:  central densities in the 059 core rise from 108 to 200 cm$^{-3}$ and from 
1600 to 5000 cm$^{-3}$ in the 073 halo.  Photodissociation is unlikely to delay collapse in the 
first two halos because they evolve little even in the absence of LW flux.  H$_2$ in the cores 
the denser halos will likely be shielded in some cases, allowing densities to continue to rise 
even as the front evaporates them.  The evolution time scales of these profiles are consistent
with free-fall times t$_{ff} = (3 \pi/32G\rho)^{1/2}$ at the halo centers, which vary from 1.17 
Myr to 38.2 Myr.  Significant contraction would be expected in the densest structure over 5 Myr
while very little would be expected to occur in the most diffuse halo, as corroborated by our
findings.

\section{Results}

We find four outcomes for the halos: (1) undisturbed cores; (2) complete disruption by an R-type 
or D-type front; (3) perturbed cores, probably with accelerated star formation; (4) a deformed 
core partially exposed to the IGM and eroded over time by ionized outflows.  The photoionization
of the TIS halo described in $\S$ serves as a rough template for these models.

\subsection{Undisturbed Cores}

Relatively dense halos are marginally ionized by the central star, and the shock driven by the 
I-front into the cloud mostly dissipates before reaching the core.  Molecular hydrogen in the 
cores of these halos is shielded from the Lyman-Werner flux of the star, even without the H$_2$ 
formed in the front.  Core collapse thus continues as the front ionizes the outer layers 
of the halo.  In Fig \ref{fig:073500pcpr} we show profiles of ionized fraction, 
density, temperature, pressure, H$_2$ fraction, and velocity along the $z$-axis at three stages 
of photoevaporation in the 073\_500pc run, with n$_c$ = 1596 cm$^{-3}$ and the star 500 pc from
the center of the halo.  
At 200 kyr the ionization front is still R-type (it becomes D-type 50 pc from the center of the 
cloud).  Lyman-Werner photons passing through the front partially dissociate 
the outer halo but the core remains deeply shielded and molecular hydrogen levels rise rapidly 
there.  Panel (a) of Fig \ref{fig:073500pcH2} shows that the core blocks LW photons in a $\sim$ 
10 pc band centered on the $z$-axis.  Central molecular hydrogen fractions rise from 2 $\times$ 
10$^{-6}$ to 1 $\times$ 10$^{-4}$ before the front becomes D-type, leveling off thereafter for 
the lifetime of the star.  H$_2$ fractions in the R-type front climb to 2 $\times$ 10$^{-4}$ and 
remain steady after the front transforms to D-type.

The D-type front again has a cometary appearance but axial compression of the shadow is much slower 
due to the greater densities in the outer regions of the cloud.  The cylindrical shock never reaches 
the axis in this run.  Perturbations are present in the front above the axis at 800 kyr.  These 
fluctuations may be an early stage of instability that 
arises in D-type fronts in which UV photons are oblique to the shock \citep{rjw02}.  The 
perturbations are longer further from the central axis where radiation is incident to the 
front at smaller angles.  LW photons preferentially stream through the underdensities, dissociating 
H$_2$ beyond in bands visible between z = 50 pc and z = 100 pc at 2.29 Myr in panel (b) of Fig 
\ref{fig:073500pcH2}.  These features are transient, vanishing by the time the star dies.  The 
shock decelerates from 10 km/sec at 50 pc to 7.5 km/sec 25 pc from the core at 2.5 Myr.  

When recombination commences at 2.5 Myr molecular hydrogen forms most rapidly in the shock but then 
radiates outward down the density gradient in the ionized regions of the halo, as shown at 2.59 Myr 
in panel (c) of Fig \ref{fig:073500pcH2}.  H$_2$ fractions rise more gradually in the core over the 
following 2.4 Myr, by approximately 50\%.  The 5.0 Myr curves in Fig \ref{fig:073500pcpr} show
that the center of the cloud is impervious to recombination flows: the incoming shock (eventually 
approaching to within a few pc), the rarefaction wave, ionized backflows at larger radii, and the 
compression of the shadow toward the axis.  Perhaps the most intesting feature of the relic H II 
region is the appearance of Rayleigh-Taylor instabilities at the interface of the warm ionized gas 
and shock remnant in panels (b) and (d) in Fig \ref{fig:073500pcrhoT}.  The fingers of partially 
ionized (and rapidly cooling) gas lengthen along the arc away from the axis, but do not affect
the dynamics of the core.

Cooling continues in the center of the halo throughout the simulation, with core temperatures falling 
from 200 K to 100 K and densities rising from 1596 cm$^{-3}$ to $\sim$ 3000 cm$^{-3}$ over 5 Myr.  The 
shock remnant eventually ripples through the center but with only minor density fluctuations.  Radiation
does not deter star formation in this model.  The evolution of the halo 1000 pc from the star is basically 
the same and radiation does not interrupt the collapse of the core.  We find in both cases that the core
continues to contract while the front photoionizes the halo, reaching final densities similar to those
when no star is present (see Fig \ref{fig:norad}).

\subsection{Complete Core Disruption}

In the other extreme, diffuse halos are easily destroyed by the radiation of the star wherever they
reside in the cluster, either by an R-type front that ionizes the cloud on time scales much shorter
than its dynamical time or by a D-type front that snowplows gas from the core at speeds greater than
its escape velocity.  We show in Fig \ref{fig:023500pcpr} the evaporation of the 023 halo (n$_c$ = 
1.43 cm$^{-3}$) along the z-zxis 500 pc from the star.  The star photodissociates the cloud well 
before the front reaches it and the equilibrium H$_2$ fractions drop to less than 1\% of their 
original value.  Molecular hydrogen fractions in the front rise over time, from 2 $\times$ 10$^{-5}$ at 225 
kyr to 2 $\times$ 10$^{-3}$ at 2.5 Myr.  However, the relatively low densities in this halo profile 
cannot produce sufficient H$_2$ to shield the core so its levels remain too small to provide significant 
cooling. The R-type front 
executes a transition to D-type $\sim$ 15 pc from the core.  The halo casts a narrow shadow (panels 
(a) and (c) of Fig \ref{fig:023500pcrhoT} that is crushed into the $z$-axis before the death of the 
star. The shock heats to over 15,000 K, inciting rapid molecular hydrogen catalysis along several 
segments of the axis at once by collisional ionization as shown in the 2.5 Myr ionization, H$_2$, 
and temperature profiles in Fig \ref{fig:023500pcpr}.  

Before the star dies the shock focuses the core into a compact fragment whose densities at 2.5 Myr 
exceed 400 cm$^{-3}$.  The shock drives this clump 25 pc from the center of the halo by 2.5 Myr at 
velocities of 10 - 20 km/sec.  As gas recombines it recedes from both the fragment and the axis in
spherical and cylindrical contours evident in panels (b) and (d) of Fig \ref{fig:023500pcrhoT}. 
The fragment remains fairly stable along the axis but expands somewhat over time as shown in the 5.0 Myr 
density profile.  The front of the clump is shocked gas at 1000 - 1500 K but it quickly cools to a 
few hundred K due to H$_2$ fractions of 5 $\times$ 10$^{-4}$ to 1 $\times$ 10$^{-3}$.

Can the clump fragment into a star?  At the fragment's average density and temperature of 3 cm$^{-3}$ 
and 150 K, its Jeans mass
\begin{equation}
m_J = \displaystyle\left(\frac{5k_BT}{G \mu m_H}\right)^{3/2}\;\displaystyle\left(\frac{3}{4\pi\rho_{cl}}\right)^{1/2},
\end{equation}
is 7.4 $\times$ 10$^{4}$ M$_\odot$, far greater than the $\sim$ 20 M$_\odot$ in the clump.  This 
together with its expansion from 300 cm$^{-3}$ to 3 cm$^{-3}$ over 2.4 Myr preclude its collapse into 
a star.  

The 023 halo is destroyed at all four distances from the star.  When it is only 150 pc away it is flash
ionized by an R-type front in 200 Kyr.  Pressure gradients in the now ionized but otherwise undisturbed 
density profile drive strong outflows that evict all the gas from the halo within 3 Myr at speeds far 
above the escape velocity.  These gradients persist until the original density profile is paved nearly 
flat: 4 $\times$ 10$^{-2}$ cm$^{-3}$ at 5 Myr.  The relic H II region uniformly cools to only 6000 K and 
ionized fractions of 50\% at 5 Myr because recombinations are so strongly suppressed in the diffuse gas.  
Molecular hydrogen is collisionally dissociated to extremely low levels by the R-type front but reforms 
soon after the death of the star.  Final H$_2$ fractions mirror the densities at the end of the run and 
are roughly level at 1 $\times$ 10$^{-4}$. 

The I-front becomes D-type at z = 0 pc, the very center of the halo, at 220 kyr when the halo is 250 pc 
from the star.  As at 500 pc, the cloud is first strongly dissociated by LW flux and the I-front shock 
again forges a small, cool dense fragment on the axis that survives to 5 Myr, but it is less massive and 
more turbulent.  With densities $<$ 1 cm$^{-3}$, the fragment is swept from the halo by the shock at 
speeds of 5 - 7 km/sec.  These lower ejection velocities are governed by the I-front driving the shock
rather than the stronger pressure gradients of the fully ionized isothermal halo in the 023\_150pc model.
When the halo is 1000 pc from the star the front becomes D-type 25 pc from the core, which is dissociated 
to H$_2$ fractions of 1 $\times$ 10$^{-7}$.  A denser clump is formed than in the 023\_500pc model, but its 
mass again falls far short of the Jeans mass.  We note that strong rarefaction waves form and evacuate gas 
exterior to the clump from the halo; they are visible as the green circular arc in panel (b) of Fig 
\ref{fig:023500pcrhoT}.  They are in all 023 runs in which the front becomes D-type.  As a rule, complete 
collisional or LW dissociation of the halo occurs whenever the core is totally disrupted by the front.

The evolution of the 039 halo 150 pc and 250 pc from the star closely resembles that of the 023\_250pc, 
023\_500pc, and 023\_1000pc models.  The fragment formed in the 039\_250pc run was 2300 M$_\odot$,
the most massive encountered in the completely disrupted halos.  This mass approaches the Jeans mass for
the density of the clump ($\sim$ 400 cm$^{-3}$) raising the interesting possibility of cloud fragmentation
into a new star in slightly denser halos illuminated at somewhat greater distances.  Had HD cooling been 
included in this model, the temperature of the fragment might have fallen to the CMB temperature, $\sim$ 
50 K at the redshifts of these halo profiles, reducing its Jeans mass by a factor of five and
ensuring its breakup.  Nevertheless, we conclude that radiation feedback on star formation in this sector of 
the flux-n$_c$ plane is overwhelmingly negative.  Compression of the halo into a dense fragment will be a 
general feature of diffuse halo evaporation because of the rapid destruction of their shadows, even in three 
dimensions.  Halos evolved from cosmological initial conditions exhibit roughly spherical morphologies 
that may cause similar focusing.
 
\subsection{Accelerated Collapse}

\begin{figure*}
\epsscale{1.15}
\plotone{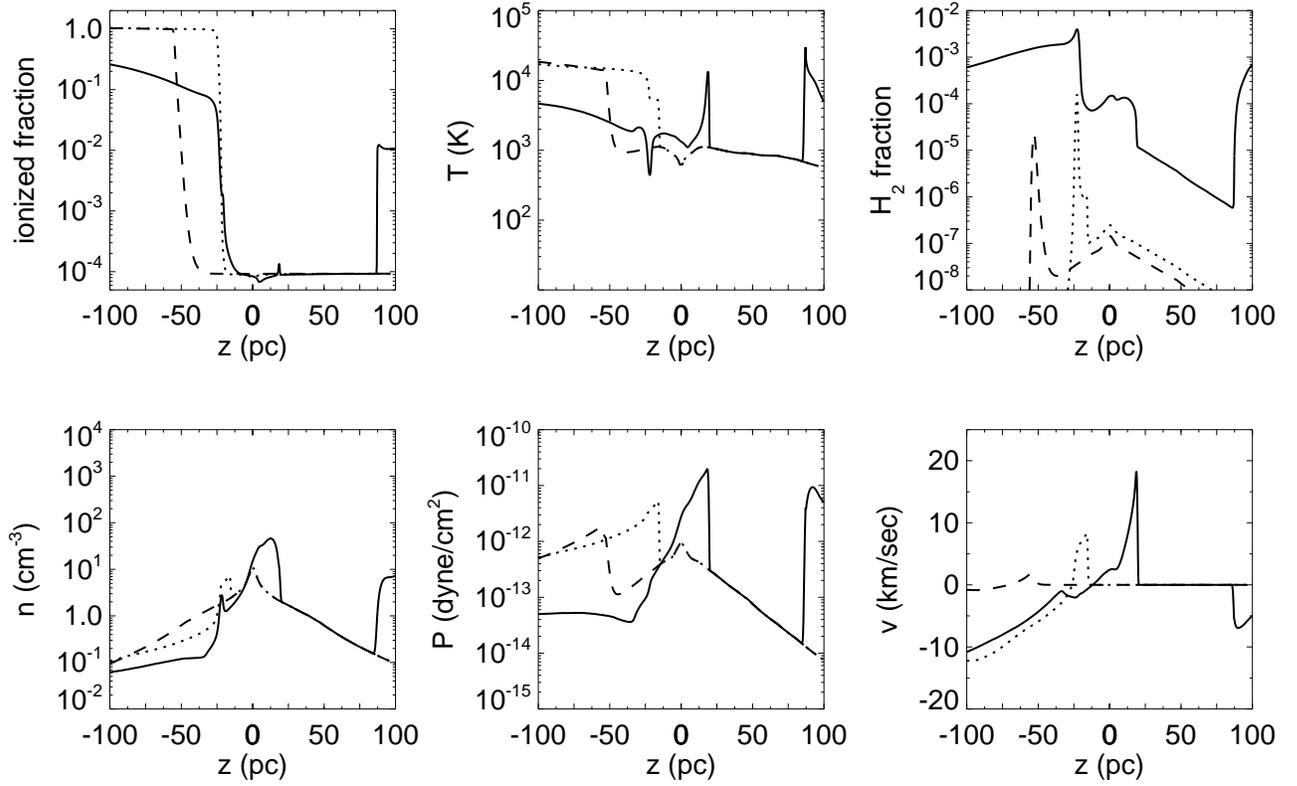}\vspace{0.15in}
\caption{Ionized fraction, density, temperature, pressure, H$_2$ fraction and velocity profiles for
the 039\_500pc model. Dashed line: 225 kyr (the R-type front), dotted line: 2.4 Myr (the D-type front), 
solid line: 5.0 Myr (the relic H II region). \label{fig:039500pcpr}} 
\vspace{0.075in}
\end{figure*}

Within a band of fluxes and central densities, the ionization front can jostle the core of the halo with a 
shock enriched with H$_2$, compressing and speeding its collapse without destroying it.  We show in Fig
\ref{fig:039500pcpr} flow profiles along the central axis for the 039 halo (n$_c$ = 10.5 cm$^{-3}$) 500 pc
from the Pop III star.  The star initially photodissociates the cloud, leaving only 5\% of the original 
H$_2$ in the core at 225 kyr.  The front converts to D-type 40 pc from the center and is 15 pc away when 
the star dies.  The velocity of the front is 10 km/sec at 2.4 Myr in Fig \ref{fig:039500pcpr} (having been
constant from -40 pc to -15 pc) but is less than 2.5 km/sec upon reaching the core at 5.0 Myr.  Gas just 
beyond the core is shot forward as the halo shadow squeezes the $z$-axis, but the gas in the core continues 
to slow.  The density at the center of the cloud is 30 cm$^{-3}$ and its molecular hydrogen fraction is 2 
$\times$ 10$^{-4}$ at 5.0 Myr.  

The gas launched forward from the halo by axial implosion is at speeds well above the escape velocity  
and cannot contribute to star formation in the core.  This jet would not be as focused in a three 
dimensional simulation with an actual halo but it would still tend to eject matter from the far side.  
The core is somewhat displaced from the center of the halo but remains there in spite of the
jet. With final densities nearly three times what would have been present at 5 Myr in the absence of 
radiation, star formation is probably accelerated in this core.  However, to capture the true time scales 
on which the 1000 K core cools and collapses into a star we must follow its displacement from the center 
of the halo with high-resolution adaptive mesh refinement (AMR) calculations currently under development.
 
\begin{figure*}
\epsscale{1.0}
\plotone{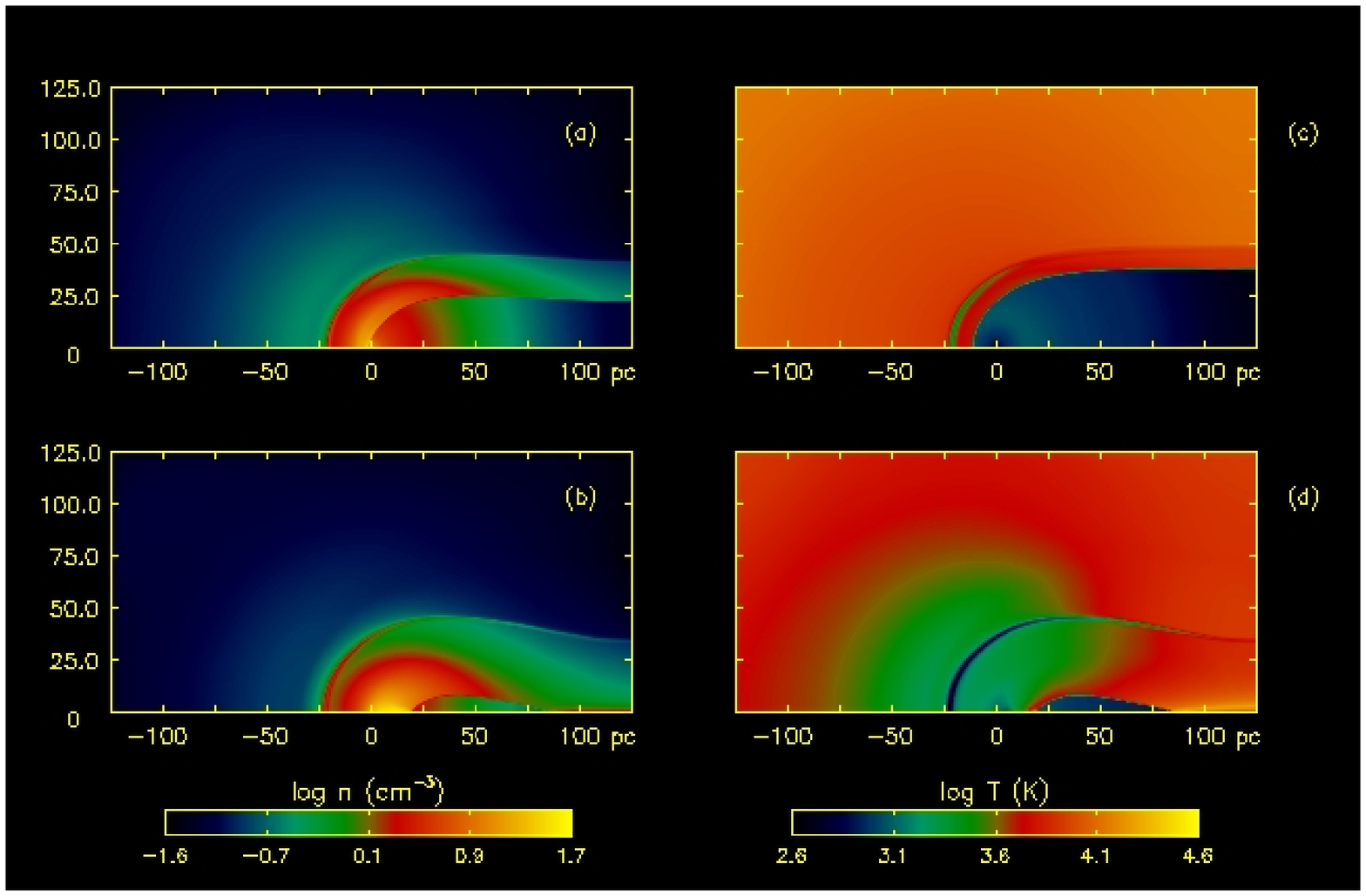}\vspace{0.15in}
\caption{Halo evaporation: model 039\_500pc.  Panels (a) and (b) are densities at 3.65 Myr and 
5.0 Myr, respectively.  Panels (c) and (d) are temperatures at 2.75 Myr and 5.0 Myr, respectively. 
\label{fig:039500pcrhoT}} 
\vspace{0.075in}
\end{figure*}

The density and temperature panels in Fig \ref{fig:039500pcrhoT} reveal two interesting features.  First,
a strong rarefaction wave again detaches from the rear of the halo in the recombinational flow, rapidly
cooling as it grows.  At 5 Myr the wave has withdrawn to z = -25 pc with a density of 2 - 3 cm$^{-3}$ and 
temperature of $\sim$ 300 K.  Second, the shadow strikes the $z$-axis at two positions at once, z = 15 pc 
and z = 75 pc.  The implosion rebounds from the axis beyond 75 pc, again creating molecular hydrogen by
 collisional ionizations.

\begin{figure*}
\epsscale{1.15}
\plotone{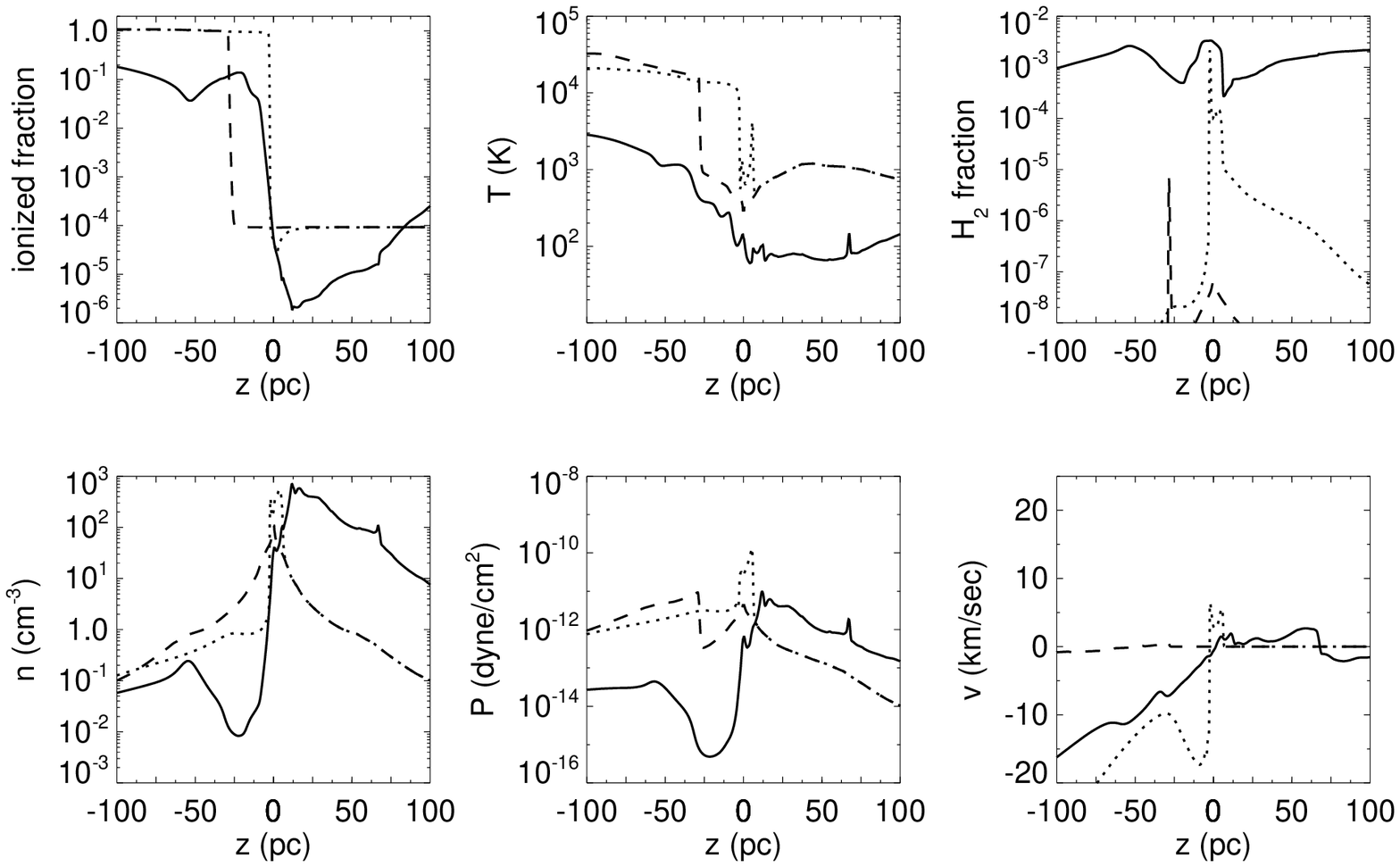}\vspace{0.15in}
\caption{Ionized fraction, density, temperature, pressure, H$_2$ fraction and velocity profiles for
the 059\_150pc model. Dashed line: 50 kyr (the R-type front), dotted line: 2.5 Myr (the D-type front), 
solid line: 5.0 Myr (the relic H II region). \label{fig:059100pcpr}} 
\vspace{0.075in}
\end{figure*}

\begin{figure*}
\epsscale{1.0}
\plotone{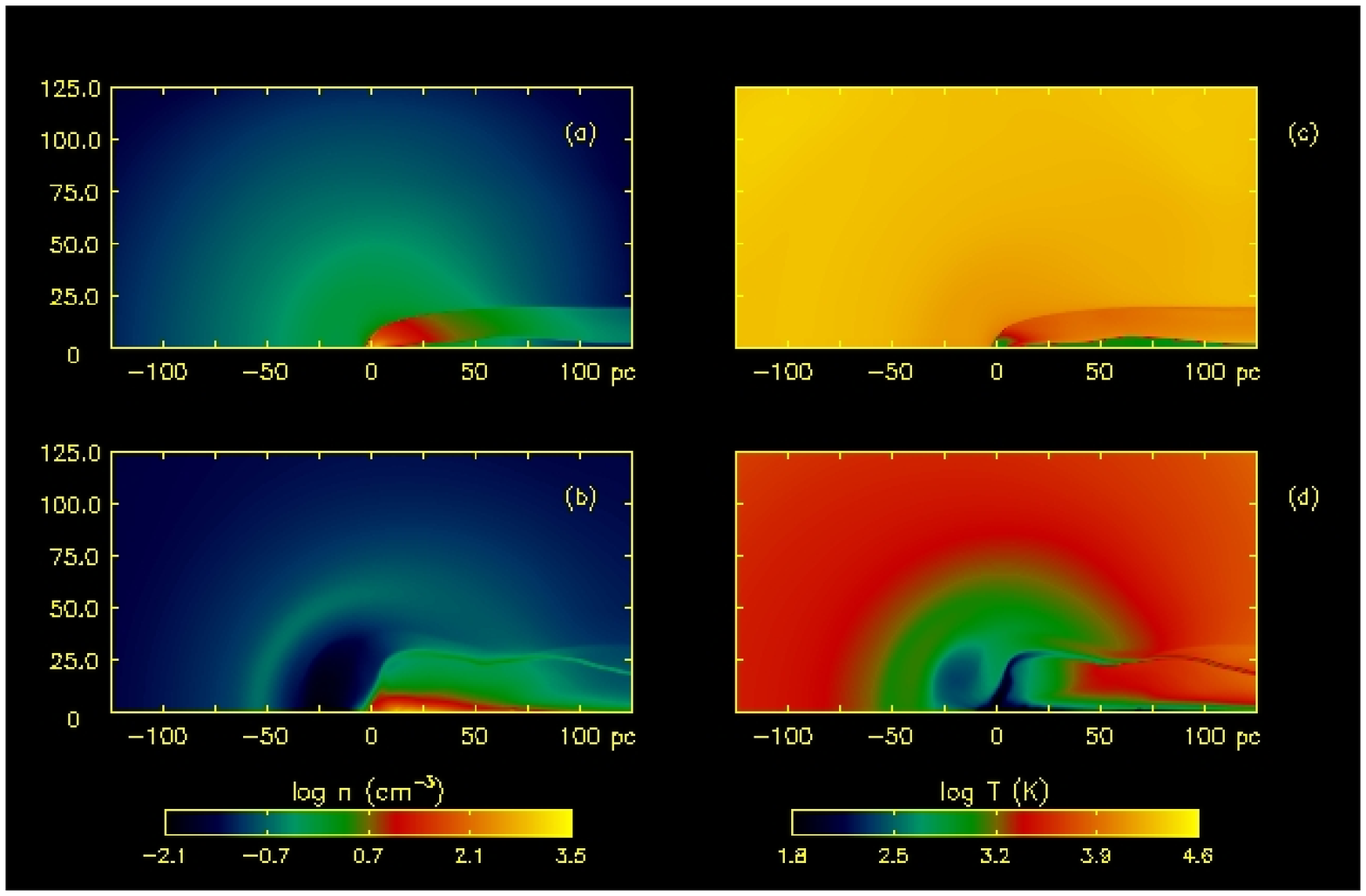}\vspace{0.15in}
\caption{Halo evaporation: model 059\_150pc.  Panels (a) and (b) are densities at 2.5 Myr and 
5.0 Myr, respectively.  Panels (c) and (d) are temperatures at 2.5 Myr and 5.0 Myr, respectively. 
\label{fig:059100pcrhoT}} 
\vspace{0.075in}
\end{figure*}

Feedback in the 039 halo 1000 pc from the star will be more conclusively positive because the remnant
slows even more before reaching the core.  Initial dissociation of the core is not as severe at this 
distance, with H$_2$ equilibrium value that rises to 2 $\times$ 10$^{-6}$ by 2.5 Myr.  The shock in the 
relic H II region is over 13 pc from the core at 5.0 Myr with a rarefation wave at -40 pc.  The core 
of the 073 halo 250 pc from the star is also density enhanced but follows a different evolutionary path.  
The shock reaches the core at 3.75 Myr at velocities less than 1 km/sec, raising its density from 1600 cm$^{-3}$ 
to 5000 cm$^{-3}$ at a temperature and H$_2$ fraction of 175 K and 2 $\times$ 10$^{-5}$, respectively.  
Collapse of the core would be strongly accelerated in this instance but the cylindrical implosion 
behind the halo drives gas back into it from the right, somewhat reducing the density of the core.  
Neglecting backwash from the shadow, which would be significantly weaker in three dimensions, we 
conclude that radiation exerts positive feedback on star formation in this model.  The halos in the 
059\_500pc and 059\_1000pc runs follow similar evolutionary tracks but in these cases the shadow is 
still far from the $z$-axis at 5 Myr, so backflows do not reach the core.

Interestingly, ionization front instabilities similar to those in the 073\_500pc model develop in the 
073 halo 250 pc from the star at 400 kyr.  In this case the unstable modes become completely nonlinear
away from the axis, extending long fingers of ionized gas nearly parallel to the arc forward into the
neutral gas.  However, these fingers are crushed downward toward the axis as the shadow narrows, 
creating colliding shock fronts that roil gas in the otherwise smooth arc.  The perturbations grow much 
larger because they form sooner than in the 073\_500pc model.  However, as before they have no impact 
on the core.  This halo is partially dissociated in the sense that equilibrium H$_2$ values are 
somewhat depressed below the levels present at 500 pc and 1000 pc.  Core molecular hydrogen abundances
fall below 1 $\times$ $10^{-4}$ so the halo ceases to condense while being ionized.  

We show in Fig \ref{fig:059100pcpr} flow profiles for the 059\_150pc model at 50 kyr, 2.5 Myr, and
5.0 Myr.  The core is at first dissociated to low molecular hydrogen fractions ($\sim$ 6 $\times$
10$^{-8}$) by the star.  The ionization front forms a shock 20 pc from the halo; as it ascends the
density gradient toward the center, its density rivals that of the core just before merging with it
at 2.0 Myr.  The shock becomes nearly opaque to Lyman-Werner flux from the star and H$_2$ catalysis
shoots rapidly upward to 2 $\times$ 10$^{-4}$ from 1 Myr to 2.5 Myr, effectively reversing the initial
photodissociation.  As the shock merges with the core its density climbs from 108 cm$^{-3}$ to 500 
cm$^{-3}$ at 2.5 Myr.  At nearly the same time the core is closed off from the right as collapse of 
the halo shadow meets the $z$-axis, as shown in panels (a) and (c) of Fig \ref{fig:059100pcrhoT}.  The 
coordinated flows drive central densities to 2000 cm$^{-3}$, mostly halting their displacement to the 
right.  At 5 Myr the core is nudged 5 pc from the origin but at velocities of less than 2 km/sec. The 
density of the core diminishes to 700 cm$^{-3}$, mostly due to eddies above the origin evident in panels 
(b) and (d) of Fig \ref{fig:059100pcrhoT}.  The center of the cloud remains fairly compact and falls to 
temperatures of $\sim$ 100 K.  Velocities are nearly flat from 0 pc to 50 pc at 5 Myr. Of all the models, 
collapse of the core into a star would be amplified most in this one.  In this instance backflow from 
the shadow into the core reinforces rather than delays collapse; even though our idealized two-dimensional 
geometry overestimates this flow, it would still be present in three dimensional models.  Collapse would 
likely be enhanced, albeit somewhat off-center in the dark matter halo. 

\subsection{Drained Cores}

\begin{figure*}
\epsscale{1.15}
\plotone{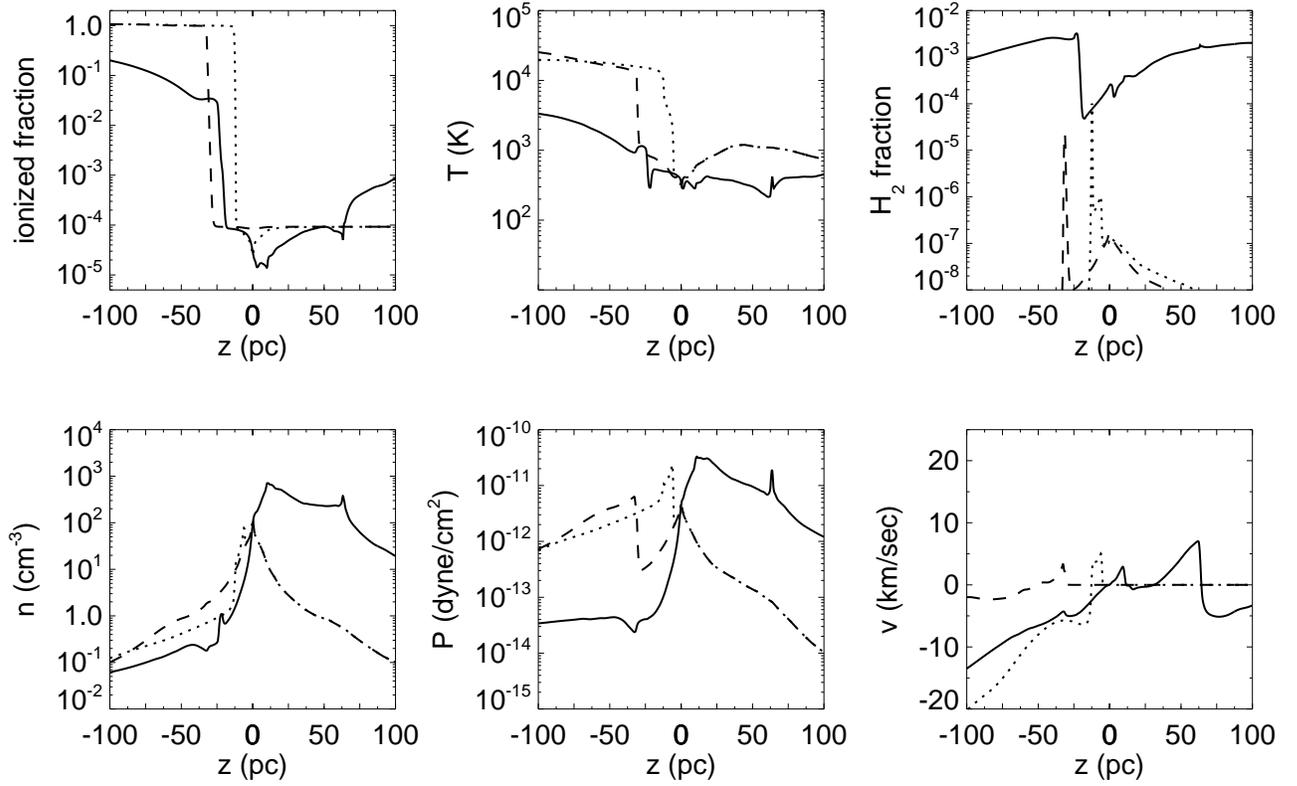}\vspace{0.15in}
\caption{Ionized fraction, density, temperature, pressure, H$_2$ fraction and velocity profiles for
the 059\_250pc model. Dashed line: 225 kyr (the R-type front), dotted line: 2.4 Myr (the D-type front), 
solid line: 5.0 Myr (the relic H II region). \label{fig:059250pcpr}} 
\vspace{0.075in}
\end{figure*}

\begin{figure*}
\epsscale{1.0}
\plotone{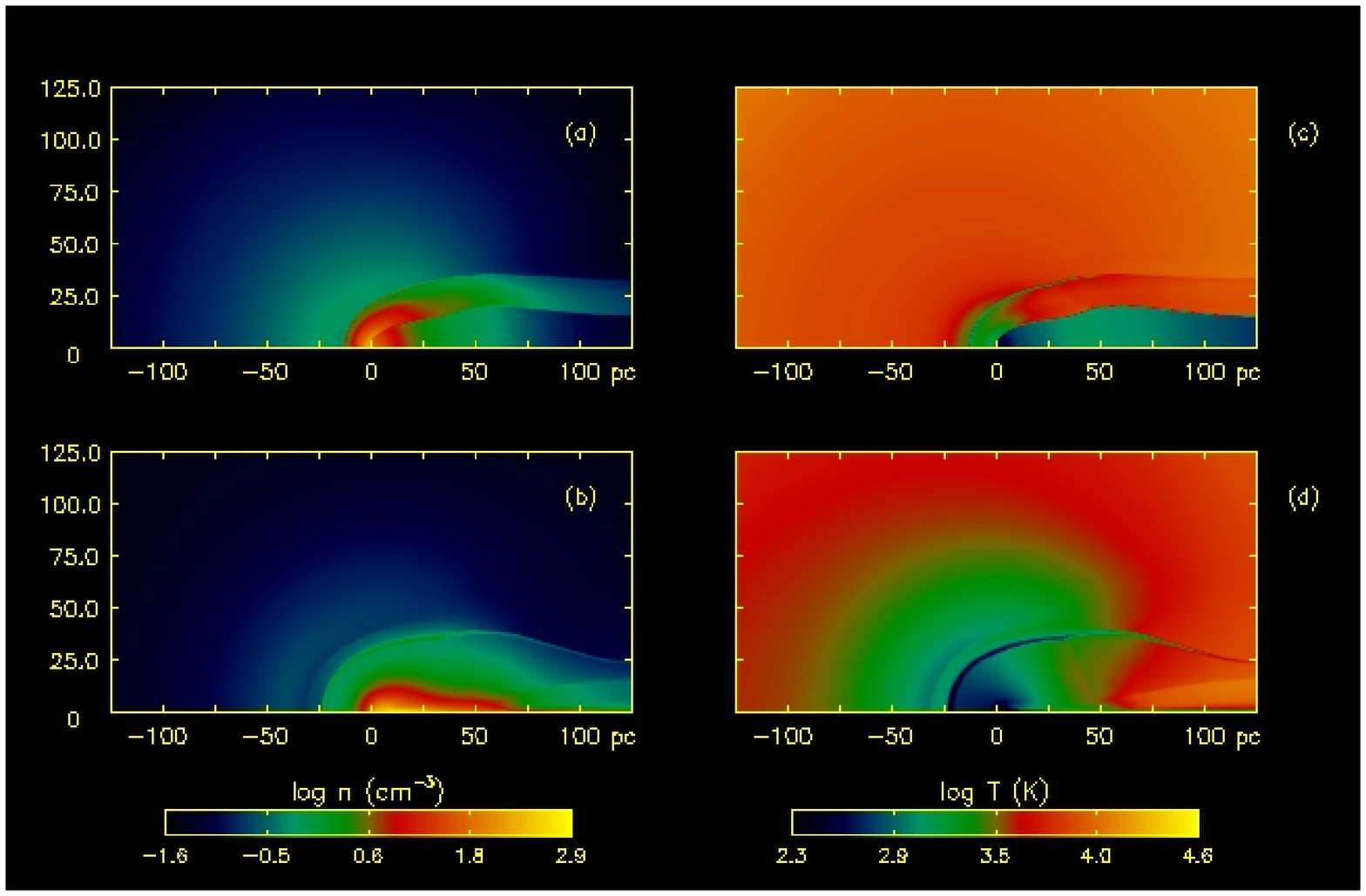}\vspace{0.15in}
\caption{Halo evaporation: model 059\_250pc.  Panels (a) and (b) are densities at 3.03 Myr and 
5.0 Myr, respectively.  Panels (c) and (d) are temperatures at 3.03 Myr and 5.0 Myr, respectively. 
\label{fig:059250pcrhoT}} 
\vspace{0.075in}
\end{figure*}

In a less forgiving sector of the flux-n$_c$ plane the shock remnant displaces the core more violently 
and the implosion of the shadow deforms it more severely, but it still survives.  In these cases the core 
tends to spring back from compression and be eroded by the rarefaction wave and relic ionized outflows.  
Central densities fall rapidly in the wake of the shock on dynamical time scales that may be shorter than 
cooling and collapse times.  Fig \ref{fig:059250pcpr} shows flow curves through the center of the 059 
halo 250 pc from the primordial star. The star dissociates the halo, reducing H$_2$ fractions at the center
to a tenth of their original value by 225 kyr.  The ionization front becomes D-type 32 pc from the core.  
Unlike the 059\_150pc run, the front never allows molecular hydrogen to reform at the center because 
at 2.5 Myr it is still 7 pc from the origin, not close enough to snowplow H$_2$-enriched gas into a shield 
capable of protecting the core from the LW photons.

The shock reaches the center of the cloud at 3.5 Myr, boosting its density from 108 cm$^{-3}$ to 300 
cm$^{-3}$.  The 3 km/sec remnant collides with a backflow shock from the collapsing shadow shortly
thereafter, driving the displaced core briefly to densities of 800 cm$^{-3}$ at 4.5 Myr.  This gas 
is moving at nearly the escape speed but will likely decelerate as it advances to the right.  However,
at z = 0 the gas is at 200 cm$^{-3}$ and moving at sub km/sec speeds.  Since it is still centered in 
the dark matter potential, one might expect its collapse to accelerate but its density continues to 
rapidly fall because of the steep velocity gradient to the left of z = 0 established when the front 
ablates the halo.  Hence, even though the core is compressed it is also smeared along the $z$-axis by 
both the remnant and the axial shock.  We cannot state the ultimate fate of the core with certainty 
in these circumstances and reserve it for further study.  The evolution of the 073 halo illuminated 
from 150 pc is very similar to that of the 059\_250pc model.  The main difference is that the D-type 
front comes close enough to the core for its shock to completely shield it from LW flux, as in the 
059\_150pc model.

\begin{figure*}
\epsscale{1.15}
\plottwo{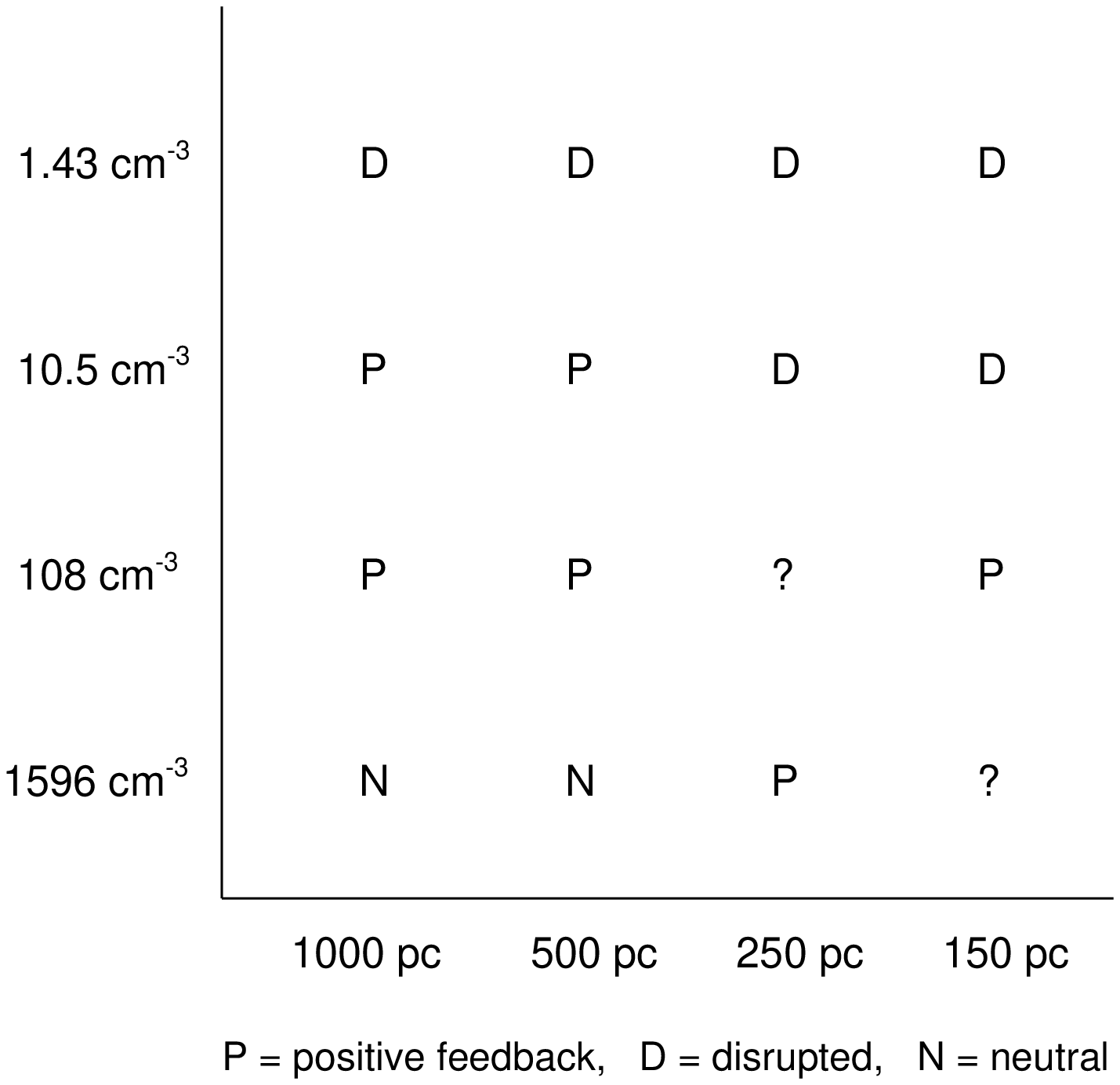}{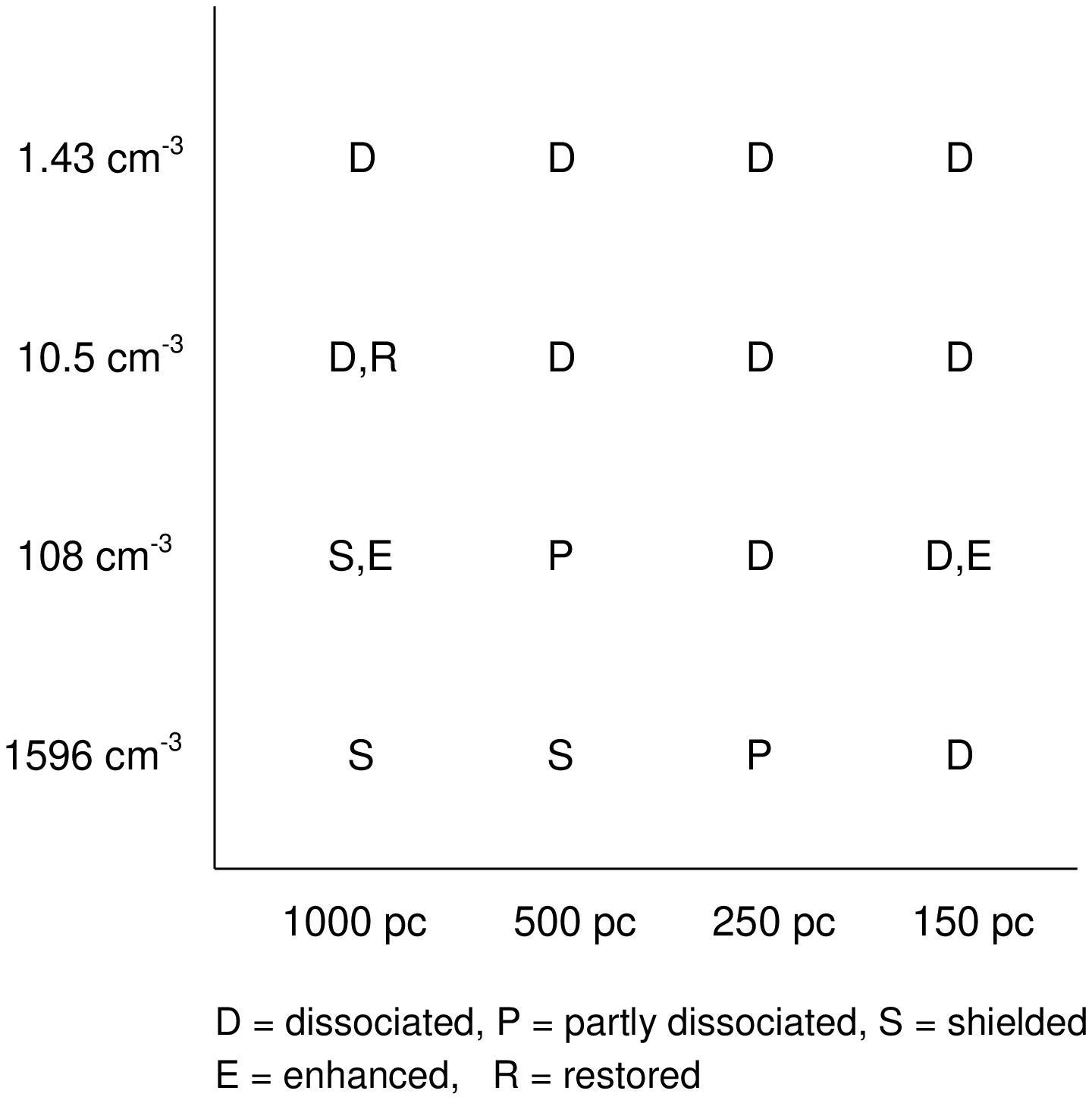}
\caption{Radiative feedback on star formation (left) and dissociation of the core of the halo (right)
in each of the photoevaporation models.  The ionizing photon fluxes at 150 pc, 250 pc, 500 pc and 1000 
pc are 5.173 $\times$ 10$^7$ cm$^{-2}$ s$^{-1}$, 1.862 $\times$ 10$^7$ cm$^{-2}$ s$^{-1}$, 4.655 
$\times$ 10$^6$ cm$^{-2}$ s$^{-1}$, and 1.164 $\times$ 10$^7$ cm$^{-2}$ s$^{-1}$, respectively.
\label{fig:SF}}  
\vspace{0.1in}
\end{figure*}

We tabulate the effect of UV radiation on star formation in each model and the degree of dissociation
of the halo prior to the death of the star in Fig \ref{fig:SF}.  A clear trend in feedback ranging
from neutral to positive to negative is evident in dense halos illuminated at large distances to 
diffuse halos close to the star.  The two question marks refer to deformed cores in which collapse is
uncertain.  The D,R entry in the dissociation table signifies that H$_2$ in the core is initially 
destroyed by the star but then allowed to reform due to shielding by the front.  The two D,E entries
are similar except that the core H$_2$ levels surpass their original values before the death of the 
star. S,E means that the original core molecular hydrogen is self-shielded from Lyman-Werner flux and 
then rises during the life of the star.  A trend in dissociation is apparent: massive cores far from 
the star retain their H$_2$ while lighter cores close to the star do not.

\subsection{Recombination Radiation}

As observed earlier, in neglecting direct transport of diffuse radiation we do not obtain the exact 
geometry of the shadow cast by the halo.  Reprocessed radiation from the ionized gas bounding the
shadow would further degrade it.  However, recombination photons above the ionization limit and in
the Lyman-Werner band could also alter the time scales of H$_2$ formation in the relic H II region. 
This effect would be transient, with the recombination luminosity greatest in the dense ionized gas
at the base of the D-type shock.  A rough estimate of its impact on H$_2$ formation can be made by 
comparison of the recombination flux and the original stellar UV flux.  The number density of all 
recombination photons emitted per unit time is
\begin{equation}
n_{rec} \; = \; n_e n_p \alpha_A(T), \hspace{0.25in} 
\end{equation}
and the recombination time scale t$_{rec}$ $=$ n$_p$/(dn$_p$/dt) $=$
\begin{equation}
t_{rec} \; = \; \frac{1}{n_e \alpha_A(T)}, 
\end{equation}
where n$_p$ and n$_e$ are the number densities of protons and electrons, respectively.  Taking the 
073\_250pc model as a fiducial case, at the base of the shock n$_p$ $=$ n$_e$ $\sim$ 90 cm$^{-3}$ 
and T $\sim$ 1 $\times$ 10$^4$ K.  From the recombination rates in Table 1 of \citet{hu94} T$_{rec}$ 
$\sim$ 850 yr and the flux of diffuse photons exiting any face of the zone is $\sim$ 4.3 $\times$
10$^8$ cm$^{-3}$ s$^{-1}$, which is briefly brighter than the number flux of the star at 250
pc, 1.9 $\times$ 10$^7$ cm$^{-3}$ s$^{-1}$.  Recombination photons from the shock can outshine the
original star but for less than 1 kyr.  This flux would briefly postpone H$_2$ formation in the vicinity
of the remnant but with little final influence on collapse of the core.  Recombination times are
much longer in the diffuse outer regions of the halo so its emissivity persists much longer but at
far lower luminosities, by factors of 1 $\times$ 10$^{-4}$ to 1 $\times$ 10$^{-6}$.  

\section{Discussion and Conclusions}

Radiation from the 120 M$_\odot$ star permits star formation in moderately evolved halos (n$_c$ $>$ 
2000 cm$^{-3}$) at distances greater than 200 pc but halts it in diffuse halos with central densities 
below 1 - 2 cm$^{-3}$ anywhere in the cluster.  Since this is likely the least massive halo able to  
host a star, radiation from the central source would probably prevent star formation in any smaller 
objects in its vicinity.  At intermediate central densities radiative feedback may be positive or 
negative depending on the flux incident on the halo.  In all cases in which the cloud is not destroyed 
outright, the star dies before the front reaches the core.  It is the velocity with which the shock 
remnant reaches the core of the halo from the left and the extent to which ionized gas in the relic H II 
region envelopes and squeezes the core from above and below that governs whether star formation therein 
is promoted or suppressed in these two-dimensional simulations.  In general, when the geometry of the 
halo shadow conspires with the ablation
remnant to compress the core of the cloud without displacing it too far from the center, star formation  
is accelerated.  Collapse of the center into a star is less certain if the shadow instead crushes the core 
along the $z$-axis and it is exposed to strong ionized backflows.  In these instances we must resort to 
detailed AMR calculations to determine the fate of the cloud.

We find in the end that Lyman-Werner radiation from the star exerts little feedback on star formation in the cluster.  Diffuse halos
are completely ionized, rendering photodissociation irrelevant, while molecular hydrogen in the cores of more
evolved structures is deeply shielded from LW flux.  Halos of intermediate densities are photodissociated
by the star but H$_2$ cooling causes little change in these structures over 2.5 Myr, even in the absence of 
radiation.  After the death of the star the H- channel rapidly restores the dissociated H$_2$ in the core, 
with little (if any) delay in its collapse.  In some cases the front itself shields the core, permitting 
H$_2$ fractions to rise beyond their original values before the death of the star.  Halos in which H$_2$ 
cooling makes any difference over 5 Myr have much higher central densities than those in this study and
would be completely shielded from LW flux. 

This is true only of the first few generations of stars before a steady LW background has built up and 
dissociating photons are present only during the life of the central star.  If successive star formation 
instead proceeds in the cluster or the halos form at later redshifts they will be illuminated from many 
lines of sight for greater periods of time.  H$_2$ formation would then be suppressed for longer intervals
in larger volumes of the halos than in our study.  However, \citet{met01} have found the effect of this 
background is to postpone (not prevent) collapse of the baryons in the halo into a star.  In this scenario
direct LW flux from the star will dominate the background until later redshifts due to the proximity of
halos in the cluster.  Each satellite would therefore be exposed to both an intermittent and a slowly
varying flux.  Unless more than one star forms in the cluster the ionizing radiation would always be well
represented by a plane wave since primordial H II regions are limited to radii of no more than a few kpc,
well below the usual separation between clusters.  This is generally true of primordial star formation since
the epoch of overlap between cosmological H II regions is at much lower redshifts. 

Upgrades to the ZEUS-MP reaction network are in progress to include HD chemistry and cooling, which can play
a significant role in the the relic H II region wherever significant H$_2$ fractions form \citep{jet07,yet07}. 
HD in the I-front and halo core can in principle cool the center of the cloud down to the CMB, possibly causing
its collapse into a new, less massive primordial star.  Primordial star formation may therefore bifurcate 
directly into a lower-mass branch in the second generation, unaided by metals ejected from supernovae.  HD 
cooling may also alter the dynamics of the shock by radiatively cooling it at the expense of its own 
kinetic energy, delaying its arrival at the core and modifying its final density and structure. 

Since we have adopted the least massive halo in which a star can form, our results can be taken as an 
upper limit for radiative feedback within the cluster because larger halos would be less affected.  There 
is considerable degeneracy in photoevaporation in the flux-n$_c$ plane so even though more massive stars
were not considered in this survey, many of our results would be relevant to brighter sources at greater 
distances from the halos.  Likewise, cores whose survival is nominal in this survey would almost certainly 
be destroyed by more luminous stars.  However, our study should be extended to less massive stars (40 - 80 
M$_\odot$) for two reasons.  First, these stars irradiate neighboring halos for longer times, and the I-front 
shock and implosion of the shadow may coordinate the compression of the core differently than more intense 
fluxes.  Second, the spectra of these stars are softer, modifying the structure and chemistry of the ionization 
front itself.  It is unclear whether more or less H$_2$ is manufactured in these fronts; although fewer hard 
photons ionize its outer layers, less LW flux dissociates the molecules, raising the question of how much 
H$_2$ the shock would deliver to the core.  
 
These results are in general agreement with \citet{su06} and \citet{s07} but differ significantly from
\citet{as07}.  First, we find that shock-induced molecule formation plays no role in star formation in 
cosmological minihalos at high redshifts because neither the D-type front nor its remnant ever heat the 
core above a few thousand K.  We observe shock-induced molecule formation in the implosion of the shadow 
but it never mixes with the center of the cloud and in any event is an artifact of the radial symmetry 
assumed for the halo.  We doubt this effect to be important in three-dimensional calculations in which 
the true morphology of the halos would break the symmetry of the implosion.  Furthermore, one-dimensional 
models fail to capture the shadow of the halo or its deformation of the core, which in some cases is quite 
serious.  That said, we find our results to qualitatively agree well with the two-dimensional halo
photoionization models of \citet{sir04}.  Although primordial chemistry was not included in those 
calculations, they exhibit similar shadowing of the I-front, transformation of the I-front from R-type to 
D-type along the central axis, and morphologies in outflow.  Even without H and He chemistry these models
much better represent the ionization of cosmological halos than one-dimensional calculations. 

However, two-dimensional models suffer from their own limitations, one being that they do not 
fully reproduce how the relic H II region envelops and compresses the evaporated halo.  The ablation 
remnant from the left together with relic H II surrounding the halo mold the core in concert; shocked 
flows from the shadow will be less coordinated or symmetric in three dimensions with real halos.  Minihalo 
photoevaporation must be revisited in three dimensions in order to assess the true role of shadow implosion 
in core dynamics.  We also note that our calculations do not address filamentary inflows into the halos,
but we do not expect this to be an important limitation.  The filaments threading through halos in our AMR 
simulations typically have overdensities no greater than fifty times the cosmic mean and so are still 
relatively diffuse in comparison to the halos into which they flow.  Such structures might briefly shadow 
I-fronts engulfing them but would be quickly photoevaporated, with little dynamical effect on the halo. It 
is safe to assume that any further inflow to the halo would be cut off in such circumstances.

Nevertheless, we maintain that these models yield accurate measures of the penetration of ionizing 
and and Lyman-Werner UV into minihalos at high redshift because the average density gradients they 
encounter are derived from cosmological initial conditions.  Our study unambiguously identifies the 
nature of the radiative feedback in many cases while singling out those in which star formation is 
uncertain for further examination in three dimensions.  Extraction of a three-dimensional halo from an 
Enzo simulation for photoevaporation on a cartesian mesh in ZEUS-MP would yield an accurately processed 
cloud core at the end of the main sequence lifetime of the star.  Reinsertion of the evaporated halo 
into Enzo for further evolution with AMR resolution, dark matter and gas dynamics, and primordial 
chemistry would follow the migration of a displaced core in the dark matter potential of the halo, 
clarifying whether a star actually forms.  

We have excluded several processes that may be important.  First, turbulent velocity fields in halos may incite 
ionization front instabilities in the cloud \citep{miz05,miz06}.  Such instabilites could puncture 
the core and prevent its collapse into a star.  Also, in contrast to the steady luminosities assumed 
in this study, stellar evolution models predict time-dependent fluxes for massive primordial stars 
that could modulate photoevaporation time scales.  Moreover, if the star detonates in a Type II or 
pair-instability supernova, metal enriched ejecta may wash over the halo, a scenario recently 
explored by \citet{cr07}.  The blast could strip the halo by ram pressure or mix it with metals, 
radically altering its time and mass scales of collapse.  Finally, we have assumed that the halo is no 
longer illuminated after the death of the star, but in reality it could be irradiated by hard x-ray 
photons from a black hole remnant capable of partially ionizing the gas without heating it to 
temperatures fatal to H$_2$ formation.  

\acknowledgments

We thank the anonymous referee, whose constructive comments improved the quality of this paper.  DW 
thanks Tom Abel, Kyungjin Ahn, Greg Bryan and Alex Heger for helpful discussions concerning these 
simulations.  This work was carried out under the auspices of the National Nuclear Security 
Administration of the U.S. Department of Energy at Los Alamos National Laboratory under Contract No. 
DE-AC52-06NA25396.  The simulations were performed at SDSC and NCSA under NRAC allocation MCA98N020 
and at Los Alamos National Laboratory.

\end{document}